\documentclass{aa}  
\usepackage{natbib}
\usepackage{array,booktabs}

\bibpunct{(}{)}{;}{a}{}{,} 

\usepackage{graphicx}
\usepackage{txfonts}
\begin{document}

   \title{Chaotic diffusion of the fundamental frequencies \\in the Solar System }

   \author{Nam H. Hoang, Federico Mogavero,
          \and
          Jacques Laskar
          }
    \institute{Astronomie et Syst\`{e}mes Dynamiques, Institut de M\'{e}canique C\'{e}leste et de Calcul des \'{E}ph\'{e}m\'{e}rides \\ 
    CNRS UMR 8028, Observatoire de Paris, Universit\'{e} PSL, Sorbonne Universit\'{e}, 77 Avenue Denfert-Rochereau, 75014 Paris, France \\
    \email{nam.hoang-hoai@obspm.fr }
    }

 
\abstract{
The long-term variations in the orbit of the Earth govern the insolation on its surface and hence its climate. The use of the astronomical signal, whose imprint has been recovered in the geological records, has revolutionized the determination of the geological timescales. However, the orbital variations beyond 60 Myr cannot be reliably predicted because of the chaotic dynamics of the planetary orbits in the Solar System. Taking this dynamical uncertainty into account is necessary for a complete astronomical calibration of geological records. Our work addresses this problem with a statistical analysis of 120\,000 orbital solutions of the secular model of the Solar System ranging from 500 Myr to 5 Gyr. We obtain the marginal probability density functions of the fundamental secular frequencies using kernel density  estimation. The uncertainty of the density estimation is also obtained here in the form of confidence intervals determined by the moving block bootstrap method. The results of the secular model are shown to be in good agreement with those of the  direct integrations of a comprehensive model of the Solar System. Application of our work is illustrated on two geological data sets: the Newark-Hartford records and the Libsack core.}

   \maketitle
%

\section{Introduction}

    \cite{milankovitch1941canon} hypothesized that some of the past large climate changes on the Earth originated from the long-term variations in its orbital and rotational elements. These variations are imprinted along the stratigraphic sequences of sediments.
    Using their correlations with an orbital solution \citep{laskar2004,laskar2010,laskar2020}, some of the geological records can be dated with precision. This method, named astrochronology, has become a standard practice in the stratigraphic community and has proven to be a powerful tool for reconstructing the geological timescale \citep[e.g.,][]{gradstein2004, gradstein2012,gradstein2020}.

    The climate rhythms found in the geological records are directly related to the Earth’s precession constant and to the fundamental secular frequencies of the Solar System: the precession frequencies $(g_i)_{i=1,8}$ of the planet perihelia and the precession frequencies $(s_i)_{i=1,8}$ of their ascending nodes. The evolution of these fundamental frequencies is accurately determined up to 60 Myr \citep{laskar2004,laskar2010,laskar2020}. Beyond this limit, even with the current highest precision ephemerides, it is hopless to obtain a precise past history of the Solar System simply via numerical integration. This limit does not lie in the precision of the determination of the initial conditions but originates in the chaotic nature of the Solar System \citep{laskar1989,laskar1990,laskar2011b}. However, because the astronomical signal is recorded in the geological data, it appears to be possible to trace back the orbital variations in the geological record and thus to constrain the astronomical solutions beyond their predictability horizon \citep{olsen1999, ma2017,olsen2019}.
    Nevertheless, a deterministic view of the Solar System is no longer reliable beyond 60 Myr, and a statistical approach should be adopted.
    Geological constraints should likewise be retrieved in a statistical setting. In this spirit, a recent Bayesian Markov chain Monte Carlo (MCMC) approach has been proposed to provide geological constraints on the fundamental frequencies \citep{meyers2018}. For such a Bayesian approach to give any meaningful constraint, proper prior distributions of the fundamental secular frequencies are required, and therefore a statistical study of the orbital motion of the Solar System planets is needed. This constitutes the motivation for the present study.

    \cite{laskar2008} performed the first statistical analysis of the long-term chaotic behavior of the planetary eccentricities and inclinations in the Solar System. \cite{federico2017} reconsidered the problem from the perspective of statistical mechanics. Our study is a follow-up of \citet{laskar2008}. We study fundamental frequencies instead of orbital elements because they are more robust and are closer to the proxies that can be traced in the geological records. This study is based on the numerical integrations of 120\,000 different solutions of the averaged equations of the Solar System over 500 Myr, 40\,000 of which were integrated up to 5 Gyr. The initial conditions of the solutions are sampled closely around a reference value that is compatible with our present knowledge of planetary motion.

\section{Dynamical model}
    \subsection{Secular equations}
        We used the secular equations of motions of \citet[][and references therein]{laskar1985,laskar1990,laskar2008}. They were obtained via series expansions in planetary masses, eccentricities, and inclinations as well as through  second-order  analytical averaging over the rapidly changing mean longitudes of the planets. The expansion was truncated at the second order with respect to the masses and to degree 5 in eccentricities and inclinations. The equations include corrections from general relativity and Earth-Moon gravitational interaction. This leads to the following system of ordinary differential equations:
        \begin{equation}
        \label{eq:SecEq}
            \frac{d \omega}{dt} = \sqrt{-1}  \{ \Gamma +  \Phi_3(\omega, \bar{\omega}) + \Phi_5(\omega, \bar{\omega})  \},
        \end{equation}
            where $\omega = (z_1,\dots,z_8,\zeta_1, \dots, \zeta_8)$ with $z_k = e_k \exp(\varpi_k)$ and $\zeta_k = \sin(i_k/2) \exp(\Omega_k) $. The variable $\varpi_k$ is the longitude of the perihelion, $\Omega_k$ is the longitude of the ascending node, $e_k$ is eccentricity, and $i_k$ is inclination. The function $\Phi_3(\omega, \bar{\omega})$ and $\Phi_5(\omega, \bar{\omega})$ are the terms of degree 3 and 5. The $16 \times 16$ matrix $\Gamma$ is the linear Laplace-Lagrange system, which is slightly modified to make up for the higher-order terms in the outer Solar System. With an adapted initial condition, the secular solution is very close to the solution of direct integration over 35 Myr \citep{laskar2004}. The major advantage of the secular system over direct integration is speed. Numerically integrating averaged equations is 2000 times faster than non-averaged ones due to the much larger step size: 250 years instead of 1.8265 days. It is thus desirable to employ the secular equations to study the statistics of the Solar System. However, we also compare their predictions to those of a non-averaged comprehensive dynamical model in Sect. \ref{sec:benchmark}.
            
    \subsection{Frequency analysis} \label{sect:FA}
        We employed the frequency analysis (FA) technique proposed by \cite{laskar1988,laskar1993FA} to extract the fundamental secular frequencies from the integrated solutions. The method finds a quasi-periodic approximation $f'(t) = \sum_{k=1}^N a_k e^{i\nu_k t}$ of a function $f(t)$ over a time span interval $[0, T]$. It first finds the strongest mode, which corresponds to the maximum of the function: 
        
        \begin{equation}
            \phi (\sigma) = \langle f(t) |e^{i \sigma t} \rangle = \frac{1}{T} \int^T_{0} \chi( t ) f(t) e^{-i \sigma t} dt, 
        \end{equation}
where $\chi (t)$ is a weight function that improves the precision of the maximum determination; it was chosen to be the Hanning window filter, that is, $\chi(t) = 1 + \cos(\pi t/T )$. The next step is the Gram-Schmidt orthogonalization. The complex amplitude $a_1$ of the first frequency $\nu_1$ is calculated via the orthogonal projection of the function $f(t)$ on $e^{i\nu_1 t}$. This mode is then subtracted from the function $f(t)$ to get a new function, $f_1(t) = f(t) - a_1 e^{\nu t}$. The process is then repeated with this newly obtained function until N desired strongest modes are obtained. This technique works very well for weakly chaotic systems such as the Solar System when variables can be decomposed into quasi-periodic modes over a sufficiently short period of time. It has been proven that this algorithm converges toward the true frequencies much faster than the classical fast Fourier transform \citep{laskar2005}. Therefore, it is a good tool for studying the chaotic diffusion of the fundamental frequencies. In this work we used a routine $\texttt{naftab}$ written in the publicly available computer algebra software TRIP \citep{gastineau2011trip}, developed at IMCCE, to directly apply the frequency analysis.
        
        To extract the fundamental secular frequencies of the Solar System, we applied the FA to the proper variables $(z_i^\bullet,\zeta_i^\bullet)_{i=1,8}$ of the secular equations \citep{laskar1990}. Each fundamental frequency is obtained as the frequency of the strongest Fourier component of the corresponding proper variable. 
        To track the time evolution of the frequencies, the FA was applied with a moving interval whose sliding step was 1 Myr. The interval sizes were 10 Myr, 20 Myr, and 50 Myr.

\section{Estimation of probability density functions}
    The samples of this study consist of the secular frequencies of the astronomical solutions that were obtained by integrating the secular equations (Eq. \ref{eq:SecEq}) from very close initial conditions. Due to this initial proximity, the correlation of the solutions in the samples lasts for a long period of time but will eventually diminish. Our objective is to obtain a robust estimation of the marginal probability density functions (PDFs) from these correlated samples. In fact, this correlation is the main motivation for our use of the estimation methods in this section. Details of our samples are described in the first part of Sect. \ref{sec:main}. 
    
    We used kernel density estimation (KDE) to estimate the time-evolving marginal PDFs of the fundamental frequencies of the Solar System. In addition, the statistical uncertainty of our density estimations (i.e., PDF estimations) was measured by the moving block bootstrap (MBB) method. 
    To our knowledge, this application of MBB for the KDE of a time-evolving sample whose correlation changes over time is new. Therefore, in order to ensure the validity of the method, we carried out several tests (see Sect. \ref{sec:MBB_test}). 
    
    \subsection{Kernel density estimation}
         We chose KDE, also known as the Parzen–Rosenblatt window method, as our preferred nonparametric estimator of the PDFs of the fundamental frequencies of the Solar System because of its high convergence rate and its smoothing property \citep{rosenblatt1956,parzen1962}.       
        
         We briefly present the method here. Let $\mathbf{X} = \{ X_1, X_2, \dots, X_n\} $ be a univariate independent and identically distributed (i.i.d.) sample drawn from an unknown distribution P with density function $p(x)$. The KDE of the sample is then defined as:
        \begin{equation}
            \hat{p}_h (x | \mathbf{X} ) = \frac{1}{nh} \sum_{i=1}^n K \left( \frac{x-X_i}{h} \right),
        \end{equation}
        where $K$ is a nonnegative kernel function and $h$ is the bandwidth of the KDE. 
        The choice of bandwidth $h$ is much more important than the choice of kernel $K$, which was chosen to be the standard normal distribution in this paper. In this work we consider bandwidths of the following form:
        \begin{equation} \label{eq:bw_rot}
            BW_{\beta} = 0.9 \min \left(\hat{\sigma}, \frac{IQR}{1.34} \right) n^{-\beta},
        \end{equation}
        where $\hat{\sigma}$ is the standard deviation of the sample, IQR is its interquartile range, and $\beta$ is a constant of choice. The bandwidth with $\beta = 1/5$ corresponds to the Silverman's rule of thumb \citep{silverman1986}. The $BW_{\beta=1/5}$ is a version of the optimal choice of bandwidth for Gaussian distributed data that is  slightly modified for better adaption to non-Gaussian data. The bias error and variance error of the KDE with this bandwidth will be on the same order of magnitude. Under-smoothing, that is, choosing a smaller bandwidth, shrinks the bias so that the total error is dominated by the variance error, which can then be estimated by the bootstrap method \citep{hall1995}; the common value of $\beta$ for under-smoothing is $1/3$.
        
        When the sample is identically distributed but correlated, KDE is still valid under some mixing conditions \citep{robinson1983, hart1996}. Indeed, in the case of observations that are not too highly correlated, the dependence among the data that fall in the support of the kernel function $K$ can actually be much weaker than it is among the entire sample. This principle is known as "whitening by windowing" \citep{hart1996}. Therefore, the correlation in the sample does not invalidate the use of KDE and only impacts the variability of the estimation (see Sect. \ref{subsect:MBB}).
        With regard to the choice of bandwidth, \cite{hall1995} suggested that using the asymptotically optimal bandwidth for the independent data is still a good choice, even for some strongly dependent data sequences.
        The samples generated by a chaotic measure-preserving dynamical system (as in the case of the numerical integration of the Solar System) resemble those of mixing stochastic processes; therefore, the theory of KDE should also be applicable for dynamical systems, although the formulation might be different \citep{bosq1995, maume2006, hang2018}. 
    

    \subsection{Moving block bootstrap}
    \label{subsect:MBB}
        Since the seminal paper by \cite{efron1979}, bootstrap has become a standard resampling technique for evaluating the uncertainty of a statistical estimation. 
        Bias and variance errors of KDE with the choice of bandwidth $BW_{1/5}$ (Eq. \ref{eq:bw_rot}) are on the same order of magnitude, and hence one should either under-smooth (i.e., choose a smaller bandwidth) to minimize the bias (Hall 1995) or use an explicit bias correction with appropriate Studentizing \citep{cheng2019, calonico2018}. 
        However, naively applying the i.i.d. bootstrap procedure on dependent data could underestimate the true uncertainty because all the dependence structure would be lost just by ``scrambling'' data together \citep{hart1996}. To remedy the problem, \cite{kunsch1989} and \cite{liu1992moving} independently devised the MBB, which later became standard practice for evaluating the uncertainty in dependent data (see \citealt{kreiss2012} for a review). Although the MBB for smooth functional has been intensively studied, the literature on MBB for the KDE of dependent data is very limited. Recently, \cite{kuffner2019} formulated an optimality theory of the block bootstrap for KDE under an assumption of weak dependence. They proposed both under-smoothing and an explicit bias correction scheme to obtain the sampling distribution of the KDE. However, good tuning parameters, which are generally difficult to find if the data are from an unknown distribution, are required to provide a decent result. In this paper we propose overcoming this problem with an inductive approach: The optimal tuning parameters obtained in a known model are tested on different models and then extrapolated to the subject of our study, the Solar System. 
        \paragraph{$\mathbf{Procedure\ of\ MBB}$}
        We briefly describe the under-smooth MBB for the KDE method (a more detailed description can be found in \citealt{kuffner2019}). We suppose that $\mathbf{X} = \{ X_1, X_2, \dots, X_n\}$ is a dependent sample of a mixing process with an underlying density function $p(x)$. The KDE of the sample is $\hat{p}_h = \hat{p}_h(x | \mathbf{X}) $; the hat above a given quantity denotes its estimated value. We used MBB to estimate the distribution of $\delta(x) = \hat{p}_h(x) - p(x)$, where $x \in \boldsymbol{\Omega}$ and $\boldsymbol{\Omega}$ is the domain of interest.
        Let $l$ be an integer satisfying $1 \leq l \leq n$. Then $B_{i,l} = \{ X_i, X_{i+1}, \dots, X_{i+l-1}\}$ with  $i \in  \{1, \dots, n-l+1 \} $ denotes all the possible overlapping blocks of size $l$ of the sample. Supposing, for the sake of simplicity, that $l$ divides $n$, then $b = n/l$. The MBB samples are obtained by selecting $b$ blocks randomly with replacement from $\{B_{1,l}, \dots, B_{n-l+1, l} \}$. Serial concatenation of $b$ blocks will give $n$ bootstrap observations: $\mathbf{X}^*_l = \{B^*_{1,l}, \dots, B^*_{b,l} \}$. By choosing sufficiently large values of $l$  (preferably larger than the correlation length), the MBB sample can retain the structure of the sample dependence.
        For $k>0$, the KDE of the bootstrap sample is $\hat{p}^*_{k,l} = \hat{p}_k(x | \mathbf{X}^*_l)$ and its expectation is $\mathbb{E}[\hat{p}^*_{k,l}] = \hat{p}_k(x | B_{1,l}, \dots, B_{n-l+1, l}) $. We define 
        \begin{equation} \label{eq:delta_MBB}
            \delta^*_{k,l} (x) = \sqrt{ \frac{k}{h}} (\hat{p}^*_{k,l} -  \mathbb{E}[\hat{p}^*_{k,l}] )
        \end{equation}
        such that if $h$ is chosen properly to reduce the bias to be asymptotically negligible with respect to the stochastic variation, then the MBB distribution $P( \delta^*_{k,l} (x) |  \mathbf{X}) $ is a consistent estimator of the error distribution $P(\delta(x))$ when $h \rightarrow 0$,   $ nh \rightarrow \infty$, $k \rightarrow 0$, $lk \rightarrow \infty$, and $n/l \rightarrow \infty$. We note that if $l=1$, then $k=h$ and MBB reverts to the under-smoothing procedure for the i.i.d sample studied by \cite{hall1995}. The efficiency of this estimator depends sensitively on two tuning parameters, $l$ and $k$. We are interested in the uncertainty of the KDE, which is characterized by the confidence interval $CI_{1-\alpha} (x)$ and the confidence band $CB_{1-\alpha}$, which are defined as: 
        \begin{equation} \label{eq:CI_true}
        P (|\delta(x)| < CI_{1-\alpha} (x)) = 1- \alpha,\\
        \end{equation}
        \begin{equation}\label{eq:CB_true}
        P (|\delta(x)| < CB_{1-\alpha} \forall x \in \boldsymbol{\Omega}) = 1- \alpha,
        \end{equation}
        where $\alpha$ denote the level of uncertainty; for example, $\alpha = 0.05$ denotes $95\%$ CI.

        They can be estimated by the MBB distribution $P( \delta^*_{k,l} (x) |  \mathbf{X}) $ as: 
        \begin{align}
                  P (|\delta^*_{k,l} (x)| &< \widehat{CI}_{1-\alpha}(x) ) = 1- \alpha, \\
                  P (|\delta^*_{k,l} (x)| &< \widehat{CB}_{1-\alpha} \forall x \in \boldsymbol{\Omega}) = 1- \alpha.
        \end{align}
        In this paper we also use CB and CI without the hat overhead to denote estimated values.
        
        Our choice of the parameters of the MBB procedure, $l$ and $k$, is based on the effective sample size $n_{\text{eff}}$, defined as \citep{kass1998}:
        \begin{equation}
            n_{\text{eff}} = \frac{n}{1 + 2 \sum_{k=1}^{\infty} \rho (k)},
        \end{equation}
        where $\rho(k)$ is the sample autocorrelation of lag $k$. The block length $l$ is chosen by the sample correlation size $l_{\text{corr}}$ as
        \begin{equation}
        \label{eq:l_corr}
            l = l_{\text{corr}} :=  \frac{n}{n_{\text{eff}}},
        \end{equation}
        and the bootstrap bandwidth is parametrized as
        \begin{equation}
        \label{eq:k_bw}
            k = h(c_0 + (1-c_0)l_{\text{corr}}^{-\gamma}),
        \end{equation}
        where $\gamma$ and $c_0$ are two optimizing constants.
        The reason for this choice of parametrization is twofold. First,  when $l_{\text{corr}} \rightarrow 1$, the sample become independent, and then $k \rightarrow h$. Secondly, the rate of change of $k$ with respect to $l$ should be greater when $l_{\text{corr}}$ is small than when it is large. Therefore, when $l_{\text{corr}}\gg1$, the optimal value of $k$ should be quite stable. We also observe experimentally that the optimal value of $k$ is indeed relatively stable at around $2h$ as long as $l = l_{\text{corr}} \gg 1$. So we simply chose $\gamma=1$ and $c_0=2$. This choice of parameters turns out to be quite robust, as demonstrated by the two numerical experiments in Sect. \ref{sec:MBB_test}.
        
        The literature on KDE and MBB focuses on stationary, weakly dependent sequences. 
        The data in our case, however, are different: They are not strictly stationary; the formulation of the mixing condition might be different \citep{hang2018}; the correlation in the sample is not constant but evolving with time; finally, and most importantly, the data structure is different. Our sample units, which are the orbital solutions, are ordered based on their initial distances in phase space. The solutions evolve over time but their order remains unchanged. Therefore, statistical notions such as correlation and stationarity should be considered within this framework for the sample at fixed values of time. 
        Because of the differences presented, an optimality theory, which is not currently available, might be needed for this case. However, we assume that it would not differ significantly from the orthodox analysis and that a decent working MBB procedure could be obtained with some good choice of parameters. This is tested in the section below.
    \subsection{Numerical experiments} \label{sec:MBB_test}
    We performed the KDE-MBB procedure on two numerical experiments to ensure its validity. The double Gaussian mixture model (DGMM) with different degrees of correlation was used to calibrate the algorithm of the KDE-MBB method, that is, to calibrate and test the tuning parameters (Eqs. \ref{eq:l_corr}-\ref{eq:k_bw}). The resulting algorithm was then applied on the Fermi map as an example of a real dynamical system for assessment.
    \begin{figure}
    \resizebox{\hsize}{!}{\includegraphics{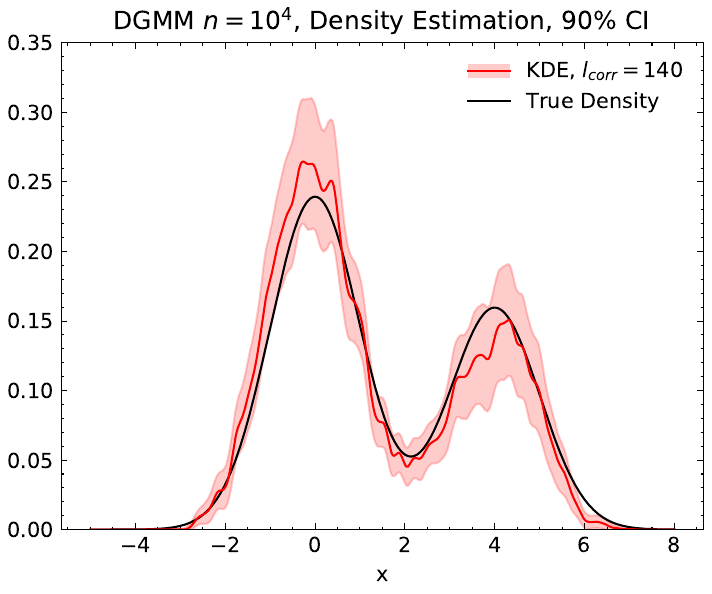}}
    \caption{ Kernel density estimation (red line) with bandwidth $h = BW_{1/3}$ and its 90\% pointwise CI (red band) from an MCMC sample of $n=10^4$ units with $l_{\text{corr}} = 140$ of a DGMM (Eq. \ref{eq:DG} - black line).}
    \label{fig:DGMM1}
    \end{figure}

    \paragraph{$\mathbf{Double\ Gaussian\ mixture\ model}$} The KDE-MBB calibration was done on MCMC sequences that sample a double Gaussian distribution. Our particular choice of the DGMM, inspired by \cite{cheng2019}, is
    \begin{equation} \label{eq:DG}
        f_{\text{DG}}(x) = 0.6\phi(x) + 0.4\phi(x-4),
    \end{equation}
    where $\phi(x)$ is the standard normal distribution. The MCMC sequence $X_1, \dots, X_n$ was obtained by the Metropolis-Hasting algorithm with a Gaussian proposal distribution whose standard deviation $\sigma_p$ characterizes the correlation length of the sequence \citep{metropolis1953,hastings1970}. We could then either vary $\sigma_p$ or perform thinning to obtain a sequence of a desired correlation length. The size of each MCMC sequence was $n = 10^4$.  The initial state of the sequence was directly sampled from the distribution $f_{\text{DG}}$ itself so that burn-in was not necessary and the whole sequence was usable.

    On each MCMC sequence, we applied the KDE-MBB procedure with 500 MBB samples and the parameters specified above (Eqs. \ref{eq:l_corr}-\ref{eq:k_bw}) with the under-smoothing bandwidth $h = BW_{1/3}$ to get the uncertainty estimation -- the standard error, the confidence interval (CI), and the confidence band (CB). Figure \ref{fig:DGMM1} shows an example of the density estimation of an MCMC sequence of $10^4$ units, with correlation length $l_{\text{corr}} = 140$; the sequence was generated by choosing $\sigma_p = 0.25$. It is clear in the figure that the true PDF (in black) lies inside the range of the 90$\%$ CI estimated by MBB.
    
\begin{figure}
    \resizebox{\hsize}{!}{\includegraphics{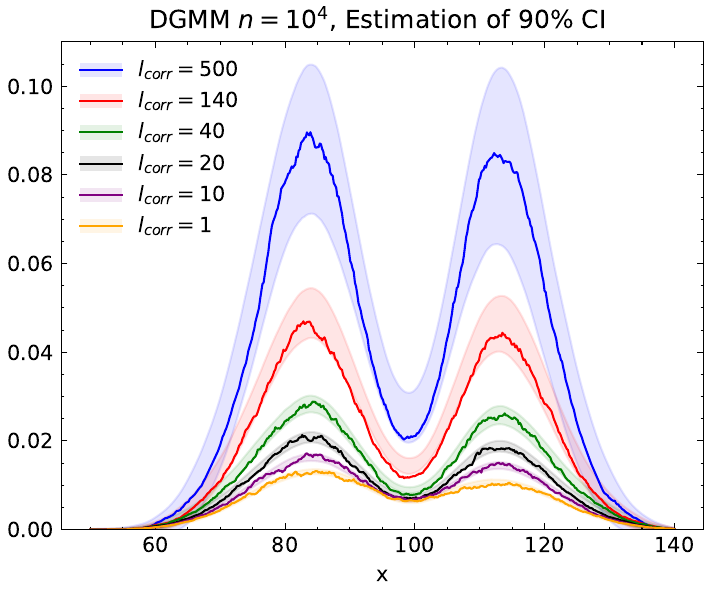}}
      \caption{Estimation of the 90\% CI of KDEs from different samples with different correlation lengths. Solid lines are the true values, and the bands denote the distribution of MBB estimations (mean $\pm$ standard deviation); both are computed from $10^3$ different samples, and each is composed of $n=10^4$ units generated from the DGMM by the MCMC algorithm with the same $l_{\text{corr}}$. }
      \label{fig:DGMM2}
    \end{figure}

    To assess the precision of the MBB uncertainty estimation, at each value of $l_{\text{corr}}$ we sampled 1000 different MCMC sequences and applied the KDE-MBB process to each of them to obtain their estimation of the CI. We thus obtained a distribution of CIs estimated by MBB, which are depicted by the bands of mean $\pm$ standard deviation  (Fig. \ref{fig:DGMM2}).
    With the knowledge of the true PDF (Eq. \ref{eq:DG}), the true CI and CB values were obtained from this collection of KDEs at each correlation length (Eqs. \ref{eq:CI_true}-\ref{eq:CB_true}).
    Figure \ref{fig:DGMM2} compares the true values of the pointwise 90\% CI with the distribution of its values estimated by MBB at various correlation lengths. 
    We see that the $90\%$ CI of KDE estimated by MBB is accurate across a very wide range of $l_{\text{corr}}$, that is, the true values almost always lie inside the estimation bands.

    \paragraph{$\mathbf{Fermi \ map}$} The second test for the MBB scheme is a two-dimensional chaotic Hamiltonian system: the Fermi map \citep{fermi1949,murray1985}. This simple toy model, originally designed to model the movement of cosmic particles, describes the motion of a ball bouncing between two walls, one of which is fixed and the other oscillating sinusoidally. The equations of the Fermi map read:
    \begin{equation}\label{eq:Fermi}
        \begin{aligned}
                u_{t+1} &= u_t + \sin \psi, \\
            \psi_{t+1} &= \psi_t + \frac{2\pi M}{u_{t+1}} \pmod {2\pi},
        \end{aligned}
    \end{equation}
        \begin{figure}
      \resizebox{\hsize}{!}{\includegraphics{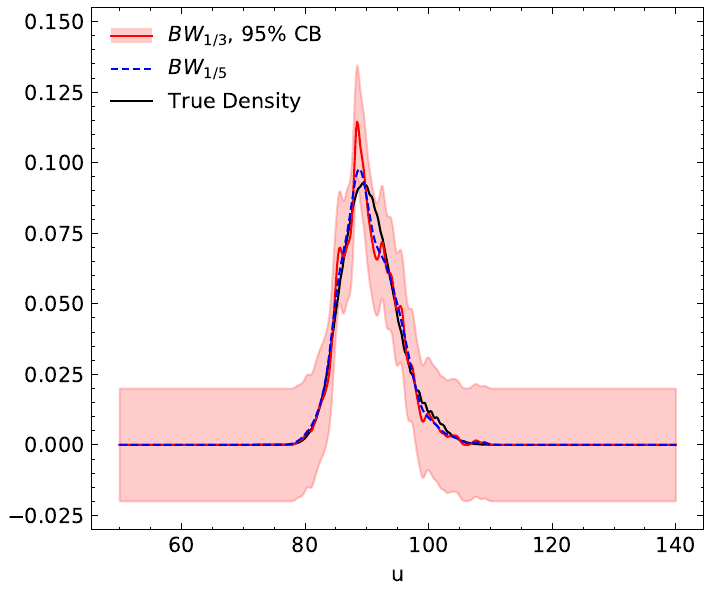}}
    \caption
      {Density estimation of Fermi map. The black line (``true'' density) is the KDE with bandwidth $h=BW_{1/5}$ from a sample of the velocity $u$ of $n=10^6$ solutions of Fermi map at t = 30. The red line denotes the KDE with bandwidth $h=BW_{1/3}$ with its 95\% CB estimated by MBB (red band) from a sample of $n=10^3$ solutions at the same time. The dashed blue line represents the KDE with bandwidth $h=BW_{1/5}$ from the same sample.}
      \label{fig:fermi1}
    \end{figure} where $u_t$ and $\psi_t$ are the normalized velocity of the ball and the phase of the moving wall right before the $t^{\text{th}}$ collision, respectively; the stochastic parameter of the system $M$ was chosen here to be $10^4$. The system was studied in the region of large-scale chaos. Our sample was obtained by evolving Eq. \ref{eq:Fermi} from $n=10^3$ initial conditions. The initial conditions, $u_0$ and $\psi_0$, are drawn from uniform distributions on $[90-10^{-5},90+10^{-5}]$ and $[0, 2\pi]$, respectively; they were then sorted by ascending value of $\psi_0$. The sorting step is imperative for quantifying the sample dependence as the initially ordered neighboring solutions still tend to be correlated afterward and the autocorrelation function can be calculated from ordered samples. Large initial phase variation will guarantee that chaotic diffusion is immediately perceptible and that the PDF of the velocity will center around its initial distribution. At each time $t$, we computed the KDE of the sample with the bandwidth $BW_{1/3}$. The KDE uncertainty was estimated by applying the MBB 400 times with the same parameters specified above (Eqs. \ref{eq:l_corr}-\ref{eq:k_bw}). The whole process was then repeated 300 times with different sets of initial conditions.
    In the Fermi map experiment, the true analytical form of the density is not available. A numerical ``true'' density, which is obtained by calculating the KDE with bandwidth $BW_{1/5}$ from an $n=10^6$ sample, was used instead to assess the validity of the MBB uncertainty estimation. From this ``true'' density and the KDEs from 300 samples, we were able to determine the true CB (Eqs. \ref{eq:CI_true}-\ref{eq:CB_true}).
    
\begin{figure}
      \resizebox{\hsize}{!}{\includegraphics{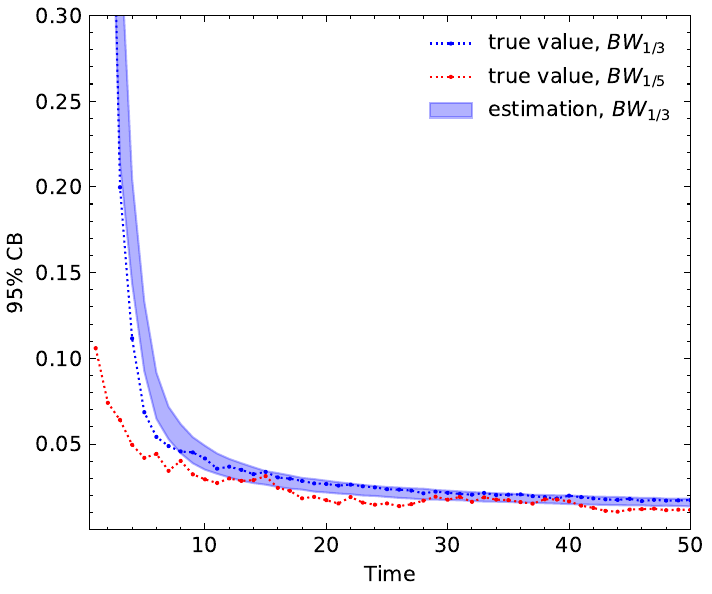}}
      \caption{Comparison of the CB estimation with its true value. The blue region represents the mean $\pm$ 3 standard deviations of the CBs estimated by the MBB method of KDEs with bandwidth $h=BW_{1/3}$ from 300 different samples, each consisting of $n=10^3$ solutions of the Fermi map; the true values of the CB are also calculated from the KDEs with bandwidth $h=BW_{1/5}$ (dotted red line) and KDEs with bandwidth $h=BW_{1/3}$ (dotted blue line) from the same 300 samples. }
      \label{fig:fermi3}
    \end{figure}

    Figure \ref{fig:fermi1} shows an example of KDE and its CB estimated by MBB at $t=30$. Although the estimated CB is valid, as the true curve lies completely in the band, the KDE itself looks quite jagged. The jagged KDE is the result of the under-smooth bandwidth $h = BW_{1/3}$, so the bias is dominated by the variance error. In fact, with the rule-of-thumb bandwidth $h = BW_{1/5}$, we can have a smoother KDE (cf. the dashed blue line in Fig. \ref{fig:fermi1}). This choice is valid because its uncertainty is always smaller than that estimated with $h= BW_{1/3}$. Therefore, we chose to use the uncertainty estimation computed with $h= BW_{1/3}$ as an upper bound of the uncertainty of a KDE with the rule-of-thumb bandwidth. For example, a $95\%$ CB of a KDE with $h=BW_{1/3}$ will represent an upper bound of a $95\%$ CB of a KDE with $h=BW_{1/5}$ (Fig. \ref{fig:fermi3}). The same also applies for the CI.
    
    Also from Fig. \ref{fig:fermi3}, we can see that the estimation of the CB follows the true value well. This is remarkable because, first, we can extend the use of MBB to measure the uncertainty of the density estimation from solutions of the chaotic dynamical system, where the correlation of the sample defined by initial distances in phase space changes over time, and second, our simple choice of MBB parameters appears to work well across very different models.

    \begin{figure*}
    \sidecaption
    \resizebox{\hsize}{!}{\includegraphics{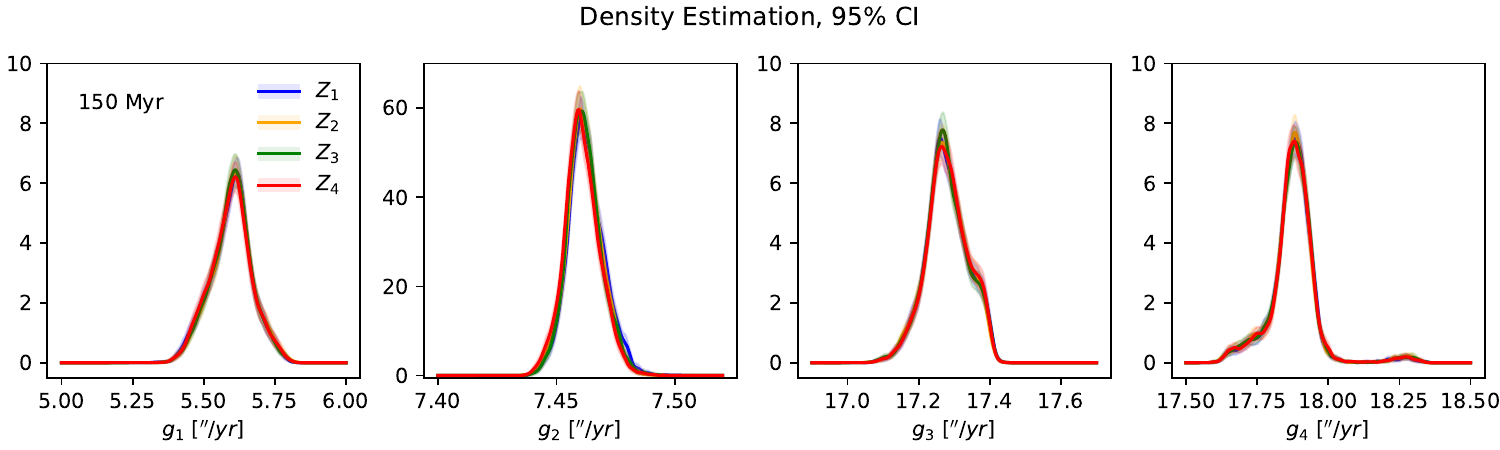}}
    \caption{Density estimation with 95\% pointwise CI of the fundamental frequencies of the four sets of the batch $\{Z_i\}_{i=1,4}$. Frequency values are obtained by FA over an interval of 20 Myr centered at 150 Myr in the future.}
    \label{fig:PDF_LZPs150}
    \end{figure*}
    
    \subsection{Combining samples}
    \label{sec:combine}
    Having obtained a well-tested uncertainty estimator of  KDE for correlated data, the practical question arose of how to efficiently combine
    the various samples of different correlation lengths $l_{\text{corr}}$. 
    Assuming we have $m$ samples of the same size n, $\mathbf{X}_1 = \{ X^1_1, X^1_2, \dots, X^1_n \} , \dots, \mathbf{X}_m = \{ X^m_1, X^m_2, \dots, X^m_n\} $, where the correlation within each sample is different, the KDE of these $m$ samples are $\hat{p}_1, \dots, \hat{p}_m$, and their pointwise standard error can be estimated by MBB as $\hat{\sigma}_1, \dots, \hat{\sigma}_m$. 
    
    If all the samples conform to the same probability density $p$, we can simply use the inverse variance weighting to get our combined KDE: 
    \begin{equation}
        \hat{p}_{\text{wa}} = \frac{\sum_i \hat{\sigma}_i^{-2} \hat{p}_i}{\sum_i \hat{\sigma}_i^{-2}}, 
    \end{equation}
    where $\hat{p}_i$ and $\hat{\sigma}_i^2$ are the individual KDE and its pointwise estimated variance, respectively. The variance of the weighted average will be $\hat{\sigma}_{\text{wa}}^2 = (\sum_i \hat{\sigma}_i^{-2})^{-1}$. With this choice, the variance of the weighted average $\hat{p}_{\text{wa}}$ will be minimized and we can thereby have the most accurate estimation of the density $p$ \citep{hartung2011}.
    
    This inverse variance weighting is only applicable if we assume that all our samples follow the same underlying distribution, as for example in the DGMM. In the case of the Solar System, different samples come from different sets of initial conditions. The densities evolving from different initial conditions will be different. Although they might converge toward a common density function after a large period of time, the assumption that different samples follow the same distribution is not generally true. However, if the differences between the density estimations from different samples are small compared to their in-sample density uncertainty, then we can assume a common true density that all the samples share, and therefore the inverse variance weighted mean and its minimized variance will in this case capture this assumed density function.    
    
    A more conservative approach consists in not trying to estimate the common true density, but rather capturing the variability stemming from different samples and combining it with their in-sample uncertainty. In this case, a combined KDE can be introduced as a pointwise random variable whose distribution function is given by:
    \begin{equation} \label{eq:comb}
        \hat{p}_{\text{c}} \sim m^{-1}\sum_{i=1}^m P( \hat{p}_i (x) - \delta^*_i (x) |  \mathbf{X}_i),
    \end{equation}
    where $\delta^*_i (x)$ is defined in Eq. (\ref{eq:delta_MBB}) and $P( \hat{p}_i (x) - \delta^*_i (x) |  \mathbf{X}_i)$ is the MBB estimation of the distribution of the pointwise KDE from sample $ \mathbf{X}_i$. If the $m$ samples are good representatives of any other sample taken from a certain population, then $\hat{p}_{\text{c}}$ is a reasonable choice for the KDE from an arbitrary sample generated from that same population.  
    We sample $\hat{p}_{\text{c}}$ in a pointwise manner. For each value of $x$, the same number of realizations are drawn from each of the $m$ MBB distributions, which are assumed here to be Gaussian for practical purposes. It should be noticed that a realization of $\hat{p}_{\text{c}}(x)$ is not a continuous curve. When needed, we could choose the pointwise median of $\hat{p}_{\text{c}}$ as the nominal continuous combined KDE.

    \section{Application to the Solar System}
    \label{sec:main}
    The goal of this paper is to have a consistent statistical description of the propagation of the dynamical uncertainty on the fundamental secular frequencies of the Solar System induced by its chaotic behavior: that is, simply put, to obtain their time-evolving marginal PDF. 
    We first sampled the initial orbital elements of the Solar System planets that were close to the reference value, and then numerically integrated the secular equations (Eqs. \ref{eq:SecEq}) from those initial conditions to obtain a sample of orbital solutions. 
    Kernel density estimation  was then used to estimate the marginal PDF of the frequencies of the sample at a fixed value of time, and finally the MBB method was applied to estimate the uncertainty of the density estimation.

    The evolution of our sample can be divided into two stages: Lyapunov divergence and chaotic diffusion. In the first stage, because the initial density is extremely localized around the reference value, all solutions essentially follow the reference trajectory; the difference between the solutions is very small but diverges exponentially with a characteristic Lyapunov exponent of $ \sim 1/5\ \text{Myr}^{-1}$ \citep{laskar1989}. The solutions in the first stage are almost indistinguishable and the correlation between them is so great that regardless of how many solutions in the sample we integrated, the effective size of the sample is close to one. 
    The second stage begins when the differences between the solutions are large enough to become macroscopically visible. The Lyapunov divergence saturates and gives place to   chaotic diffusion.
    The correlation between the solutions starts to decrease, the distribution of the sample settles, and the memory of the initial conditions fades. It can take several hundred million years for the sample to forget its initial configuration.
    Contrary to the exponential growth in the first stage, the dispersion of the samples expands slowly with a power law in time (see Fig. \ref{fig:PDF_fitParams}). The time boundary between the two stages depends on the dispersion of the initial conditions:The wider they are, the faster the second stages come, and vice versa. If they are chosen to represent the uncertainty of the current ephemeris, the second stage should take place around 60 Myr in the complete model of the Solar System \citep{laskar2010, laskar2011b}.
    
    In this section we focus on the statistical description of the fundamental frequencies of the Solar System in the second stage. The aim is to obtain a valid estimation of time-evolving PDFs of the frequencies beyond 60 Myrs. However, the PDF evolution generally depends on the choice of initial conditions. 
    Moreover, the simplification of the secular equations compared to the complete model of the Solar System could, a priori, provide results that are not sufficiently accurate. 

    \begin{figure*}
    \centering
    \includegraphics[width=18cm]{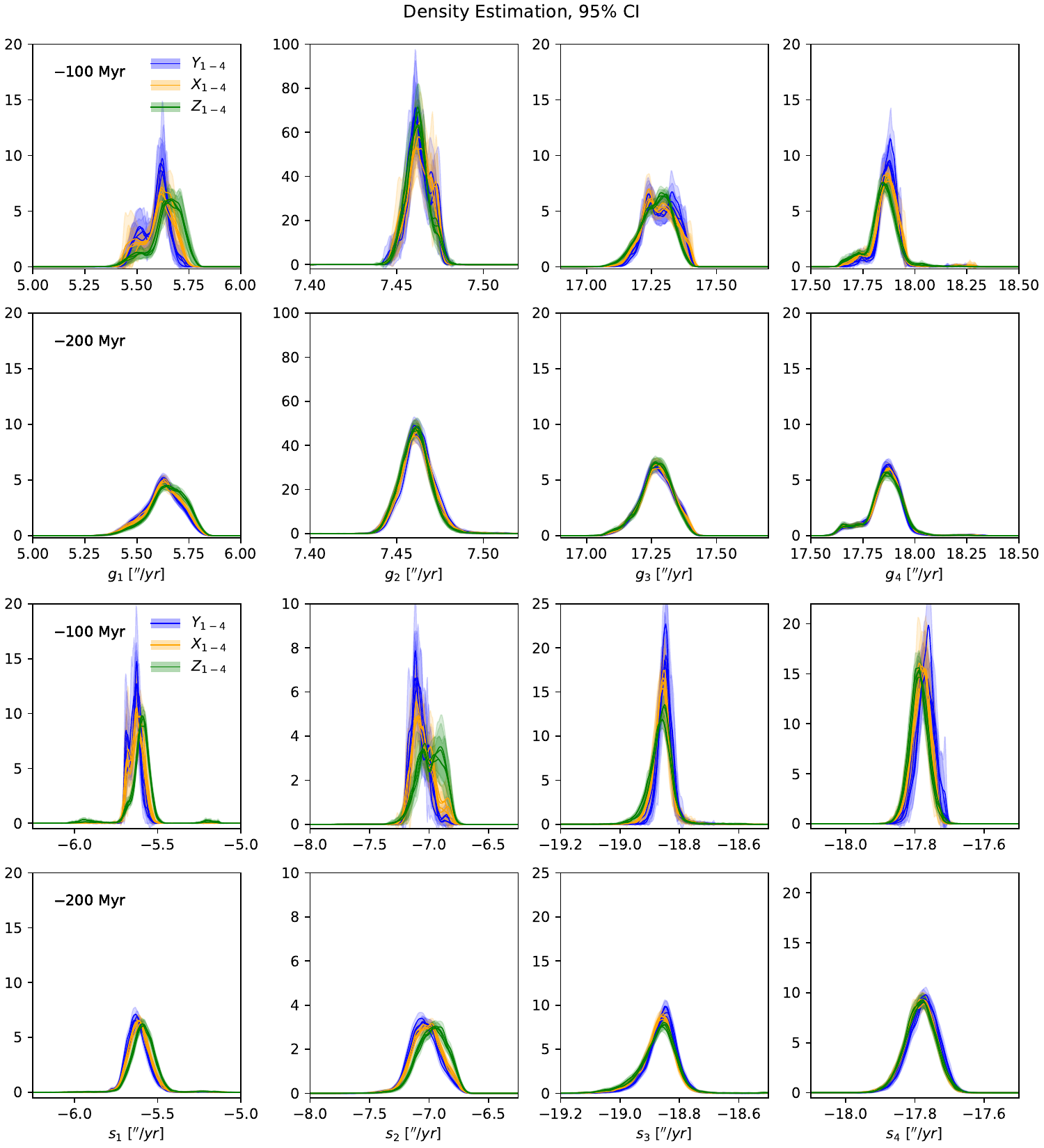}
    \caption{Density estimation with 95\% pointwise CI of the fundamental frequencies of the 12 sets coming from the three batches: $\{X_i\}$ (yellow colors), $\{Y_i\}$ (blue colors), and $\{Z_i\}$ (green colors). Estimations of sets from the same batch have the same color.  Frequencies are obtained by FA over an interval of 20 Myr centered at 100 Myr in the past (first and third row) and 200 Myr in the past (second and fourth row). The time reference is the time of $\{Z_i\}$, while the solutions of $\{X_i\}$ and $\{Y_i\}$ are shifted ahead by 30 Myr and 20 Myr, respectively. }
    \label{fig:LXYZN}
    \end{figure*}
    
    \begin{figure*}
    \centering
    \includegraphics[width=18cm] {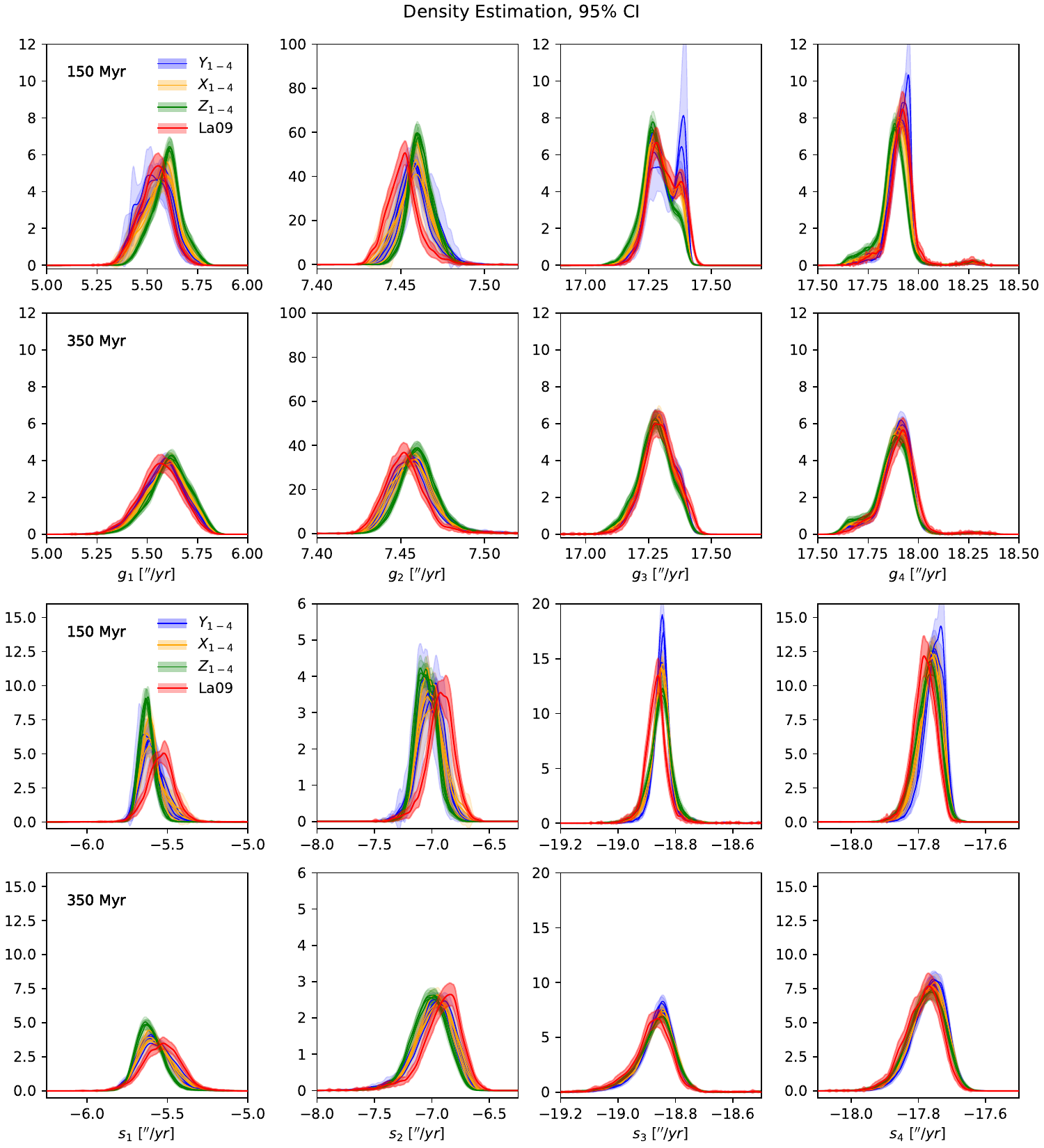}
    \caption{Density estimation with 95\% pointwise CI of the fundamental frequencies of the 12 sets coming from the three batches, $\{X_i\}$ (yellow colors), $\{Y_i\}$ (blue colors), and $\{Z_i\}$ (green colors), and from the 2\,500 solutions of complete model of \cite{Laskar2009} (red colors), which is denoted as La09. Estimations of sets from the same batch have the same color. Frequencies are obtained by FA over an interval of 20 Myr centered at 150 Myr (first and third rows) and 350 Myr (second and fourth rows) in the future. The time reference is the time of $\{Z_i\}$, while the solutions of $\{X_i\}$, $\{Y_i\}$, and La09 are shifted by 30 Myr, 20 Myr, and -70 Myr, respectively. }
    \label{fig:LXYZP}
    \end{figure*}
    \label{sec:benchmark}

        
        \subsection{Choice of initial conditions}
        \begin{table}[ht]
        
            \centering
            \begin{tabular}{c|cl}
             \hline\hline 
                Solutions & Offsets  & $\epsilon$ \\
            \hline 
                 $X_i$ & $-5000\, \epsilon$ to $5000 \, \epsilon$  &  $10^{-10}$\\ 
                 $Y_i$ & $-5000 \, \epsilon$ to $5000 \, \epsilon$  &  $10^{-8}$\\
                 $Z_i$&  $-5000 \, \epsilon$ to $5000 \,\epsilon$  &  $10^{-11}$\\
             \hline
            \end{tabular}
            \caption{Offsets of the initial eccentricities of the four planets: $\{$Mercury, Venus, Earth, Mars$\}$, which corresponds to $i=\{1,2,3,4\}$. Different integrations correspond to offsets of $N \epsilon$ in eccentricity of a single planet for $N = - 5 \times 10^3, \dots , + 5 \times 10^3$, while other variables are kept to their nominal values.}
            \label{tab:iniCon}
        \end{table}
        For a complete model of the Solar System, the initial conditions should be sampled in such a way that they are representative of the current planetary ephemeris uncertainty.
        Nevertheless, for a simplified secular model (Eq. \ref{eq:SecEq}), the difference from the complete model is greater than the ephemeris uncertainty. An optimized secular solution follows the complete solution initially but departs from it long before 60 Myr \citep{laskar2004}. A direct adaptation of the current planetary ephemeris uncertainty to the initial conditions of the secular model can thus be misleading. Therefore, we adopted a more cautious approach, that is, to study first the effect of sampling the initial conditions on the PDF estimation.
        
        Initial conditions can be sampled in many ways, especially in a high-dimensional system such as the Solar System. 
        Our choice of initial conditions is quite particular, but they encompass different possible ways of sampling initial conditions (Table  \ref{tab:iniCon}). 
        There were three batches, $\{X_i\}$, $\{Y_i\}$, and $\{Z_i\}$, which correspond to three different variation sizes $\epsilon$ of initial conditions. Each batch was composed of four different sets of samples. Each set contained 10\,000 initial conditions, where the eccentricity of the associated planet is linearly spaced from the reference value, with the spacing $\epsilon$ corresponding to the batch it belongs to.
        We then integrated the secular equations (Eqs. \ref{eq:SecEq}) from these initial conditions 500 Myr into the past and 500 Myr into the future. For the batch $\{Z_i\}$, the integration time is 5 billion years in both directions. The frequencies were then extracted using FA (Sect. \ref{sect:FA}). It should be noted that we do not aim to obtain the joint probability distribution of all the fundamental frequencies, but rather their individual marginal PDFs (i.e., the PDF of one frequency at a time).
        The marginal PDFs of the frequencies were estimated by the KDE with the rule-of-thumb bandwidth ($BW_{1/5}$); upper bounds of their 95\% CIs are measured by the MBB method with the bandwidth $h= BW_{1/3}$ and the optimized parameters (Eqs. \ref{eq:l_corr}-\ref{eq:k_bw}) from 1\,000 MBB samples.
        
        We first compare the evolution of the density of the four sets in each batch in Sect. \ref{sec:ben:samesize}, and the second test is performed to compare the statistics between the batches in Sect. \ref{sec:ben:diffsize}.  The robustness of the secular statistics is assessed by these two tests, which additionally shed light on the initial-condition-dependence aspect of the statistics.  All the density estimations from these sets are compared with those of the 2\,500 complete solutions obtained in the previous work of \citet{Laskar2009} to test the accuracy of the secular statistics. It should be recalled that this numerical experiment needed 8 million hours of CPU time, the output of which was saved and could thus be used in the present study.
        
        When comparing two sets of different sizes, because the rates of divergence in the first stage are similar, the wider set reaches the chaotic diffusion phase faster than the more compact set; hence, it is essentially diffusing ahead for a certain time in the second stage. Therefore, in order to have a relevant comparison, a proper time shift was introduced to compensate for this effect. We shifted $\{X_i\}$ and $\{Y_i\}$ ahead by 30 Myr and 20 Myr, respectively, while keeping the time of $\{Z_i\}$ as reference. This choice was motivated by the fact that the transition to the chaotic diffusion of $\{Z_i\}$ is around 50 - 60 Myr,  which is indicated by the direct integration of the Solar System \citep{laskar2010,laskar2011b}.

    \subsection{First test: Different samples of the same variation size $\epsilon$}
        \label{sec:ben:samesize}
        Comparing the density estimation of different sets in the same batch was the first test of prediction robustness from our secular model. The evolution of the density estimations are sensitive to how initial conditions are sampled, and therefore this initial-condition sensitivity must be quantified for a valid prediction. In this first test we compared the time-evolving PDFs whose initial conditions are sampled with the same variation size $\epsilon$ but in different variables. 
        
        The result is quite clear. The different sets of the same batch slowly lose the memory of their initial differences due to chaos and then converge toward the same distribution. This convergence is illustrated by Fig. \ref{fig:PDF_LZPs150}, which shows that the density estimation of $(g)_{i=1,4}$ of the four sets of the batch $\{Z_i\}$ nearly overlap with one another at 150 Myr in the future. 
        The rates of convergence of different batches are different.  
        So although $\{X_i\}$ and  $\{Y_i\}$ exhibit the same behavior as $\{Z_i\}$, they converge differently with disparate rates: At around 100-150 Myr in the future, the density estimations of the frequencies of $\{X_i\}$ nearly overlap with one another; this occurs at 150-200 Myr for $\{Y_i\}$, depending on the frequency.
        Interestingly, for the samples that are integrated in the past, the rates of convergence are higher and the overlap generally happens at around $-100$ Myr (see Fig. \ref{fig:LXYZN}).


    \subsection{Second test: Different samples of different variation sizes}        \label{sec:ben:diffsize}

        Comparing the density estimation of the three batches, $\{X_i\}$, $\{Y_i\}$, and $\{Z_i\}$, was our second test of robustness. Although the initial conditions of the three batches were varied around the same reference values, the ways they were sampled were different since the variation sizes were different. Differences in the initial variation sizes mean that the batches enter the diffusion stage at different times and also at different points in the phase space, so that the convergence between batches, if it exists, takes longer.
        The result of our test is summarized by the density estimation of the frequencies at two times in the past, $-100$ Myr and $-200$ Myr (Fig. \ref{fig:LXYZN}). At $-100$ Myr, the density estimations from different sets of each batch cluster around one another as described in the previous test. Each batch forms a cluster of density estimations, and the differences between the three clusters are noticeable. Moreover, the estimation uncertainty, depicted by the colored band, is quite large. Fast-forward 100 Myrs of chaotic mixing: at $-200$ Myr, density estimations of the frequencies spread out, estimation uncertainty shrinks, and, most importantly, differences between the three batches are much smaller and continue to diminish even further with time. In the opposite time direction, the same phenomenon is observed but the rate of convergence between the batches is slower (Fig. \ref{fig:LXYZP}). At 150 Myr in the future, the density estimations of the frequencies of $\{X_i\}$ and $\{Z_i\}$ have practically converged and those of $\{Y_i\}$ are still trying to, and yet differences between the three batches are noticeable. However, the estimations of all 12 sets from the three batches nearly overlap with one another at 350 Myr, which demonstrates that the effect of different initial samplings vanishes via chaotic mixing. It should be noted that when looking at some specific properties of the PDF, such as means and variances, the differences between the sets are small. For example, the differences between the means of the PDF estimation of the 12 sets are generally smaller than $0.1\ ''/\text{yr}$ for most of the fundamental frequencies at $-100$ Myr; at $-200$ Myr, the differences diminish to twice as small.

        \subsection{Final test: Comparison with the complete model}
        \begin{figure*}
        \centering
        \includegraphics[width=18cm] {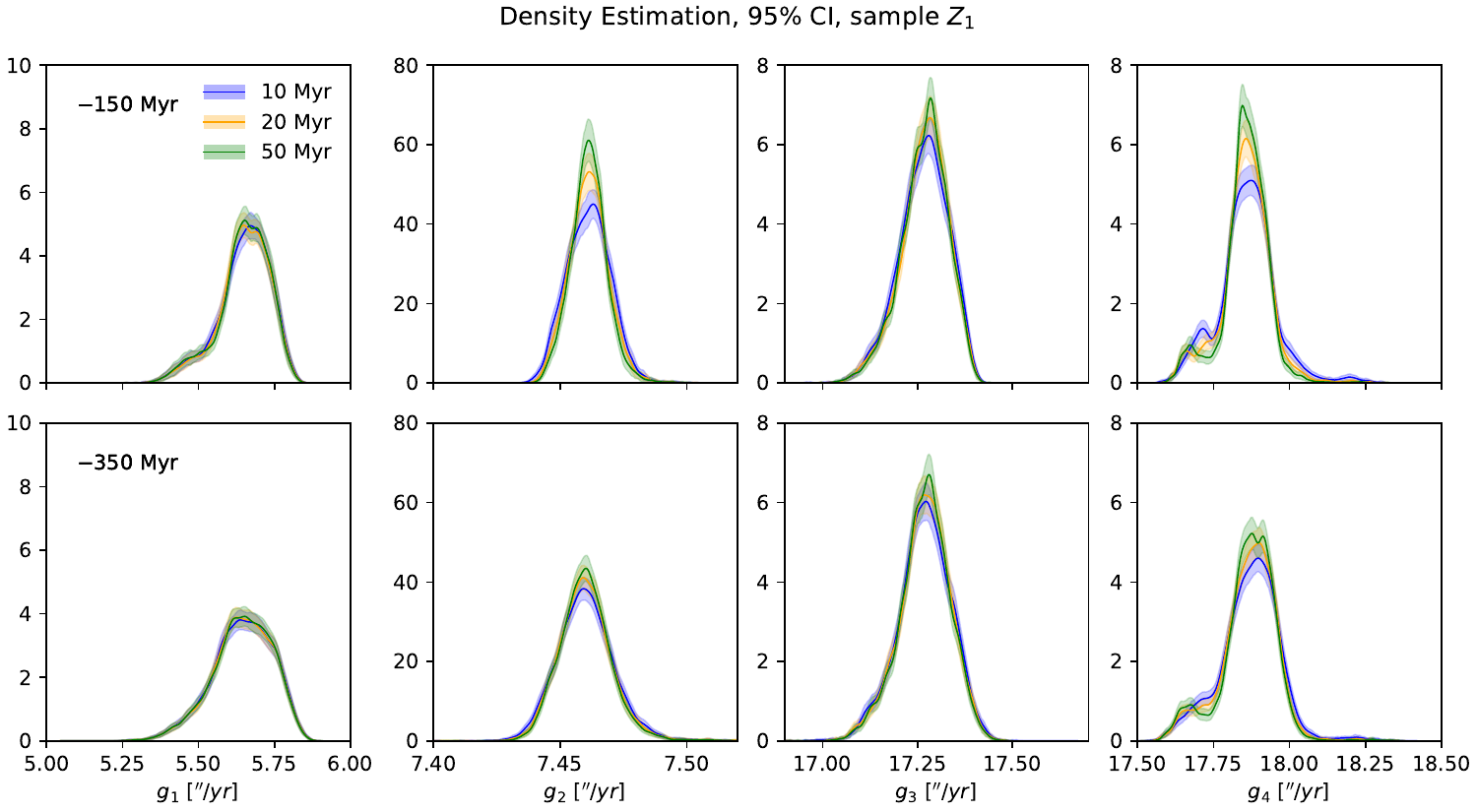}
        \caption{ Density estimation with $95\%$ pointwise CI of the fundamental frequencies of the set $Z_1$, which are obtained by FA over an interval of 10 Myr (blue colors), 20 Myr (yellow colors), and 50 Myr (green colors) centered at $-150$ Myr (top row) and $-350$ Myr (bottom row). }
        \label{fig:FA}
        \end{figure*}
           \begin{figure*}
    \includegraphics[width=18cm]{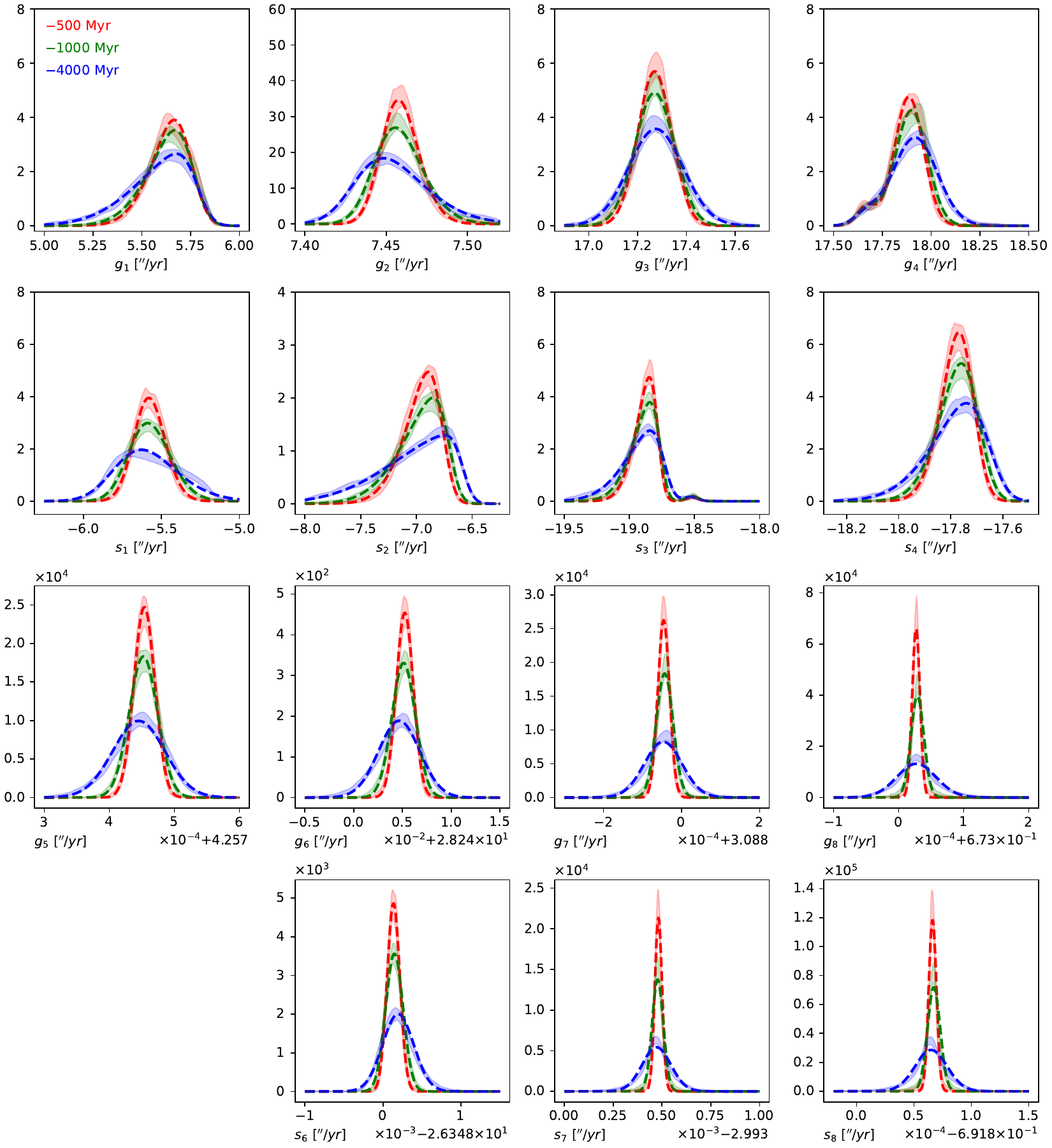}
    \caption{Density estimation with 95\% pointwise CI of the fundamental frequencies of the Solar System from the $\{ Z_i \}$ solutions in the past at 500 Myr (red band), 1 Gyr (green band), and 4 Gyr (blue band); the dashed curves with the corresponding colors denote their fitting distribution (Eqs. \ref{eq:fitModel} and Table \ref{table:params_fit}). The frequencies are obtained by FA over an interval of 20 Myr.}
    \label{fig:Pdfs_fit}
    \end{figure*}
    \begin{figure*}
    \centering
    \includegraphics[width=18cm]{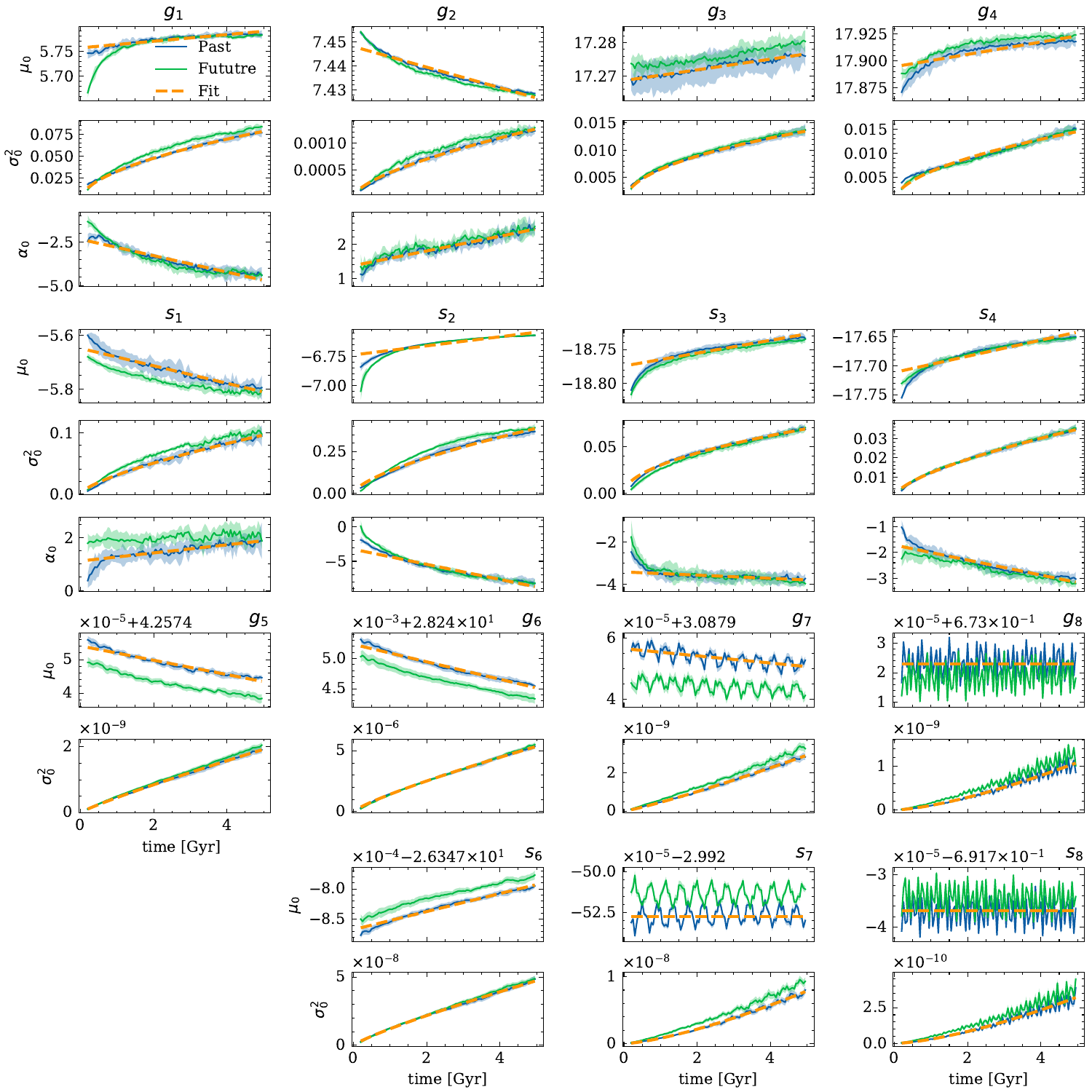}
    \caption{Evolution of the parameters of the skew Gaussian distribution in the fitting model (Eqs. \ref{eq:fitModel}) for the density estimation of the fundamental frequencies of the Solar System of $\{ Z_i \}$ solutions in the past (blue line) and in the future (green line). The bands denote the $\pm 2$ standard error. The dashed orange lines denote the power-law fits for $\sigma_0^2$ and the linear fits for $\mu_0$ and $\alpha_0$.}
    \label{fig:PDF_fitParams}
    \end{figure*}

        The convergence of the density estimations from different sets of initial conditions (seen in the first two tests) does not guarantee convergence between the secular model and the complete model of the Solar System. As a result, whether the secular statistics will resemble that of the complete model is still not clear. In this final test, we respond to this question by comparing the density estimation from our secular solutions with those obtained from the 2\,500 solutions of the complete model integrated into the future \citep{Laskar2009}. The initial conditions of the complete solutions are sampled in a way similar to the present work, that is, one variable (the semimajor axis of Mercury) is linearly spaced with a spacing of 0.38 mm and a range of about one meter. We shifted the complete solution backward by 70 Myrs to adjust for the difference in the initial variation sizes and also for the difference between the initial divergence rates. 
        
        The density estimations of the frequencies of the 2\,500 complete solutions are shown along with the estimations from the secular model in Fig. \ref{fig:LXYZP}. The estimations of the complete model are either close to or overlap with those of the secular model, even at 150 Myr. Both models even predict the same minor features in the frequency density, for example the second peaks of $g_3$ or the tails of $g_4$. The origins of these features are related to the resonances associated with $g_3$ and $g_4$, so having these same features indicates that the secular model could capture the resonance dynamics of the complete model. 
        At 350 Myr, for most of the frequencies, the differences between the results of the two models are very small, especially for some frequencies, such as $g_1$, where it is difficult to distinguish between the two models. 
        If we look at some specific properties of the PDF, such as its mean, the differences between the secular model and the complete model are on the same order as the variability of the results from the same secular model. These differences in the means of the PDFs are generally smaller than $0.05\ \arcsec/ \text{yr}$ at 350 Myr.
        Although some minor differences between the two models are still visible, especially in $s_1$ for example, these differences diminish with time. 
        This convergence between the two models strongly suggests the compatibility of the secular system with the direct integration of the realistic model of the Solar System used in \citet{Laskar2009}.

        \subsection{A complementary test on frequency analysis}
        The fundamental frequencies of the Solar System are central in our work, and the method to obtain them is thus essential. In this section we briefly examine the FA method (Sect. \ref{sect:FA}). The FA method searches a quasi-periodic approximation of the solution of the Solar System with constant frequencies over a time window $\Delta t$. So unique frequencies at time $t$ are extracted from an oscillating sequence in the time interval $[t-\Delta t/2, t+\Delta t/2]$. For a quasi-periodic solution, the longer we choose the time interval $\Delta t$ to be, the more accurate the extracted frequencies are. Nevertheless, the fundamental frequencies of the Solar System are expected to vary over a few Lyapunov times, that is, over 5 Myr. Therefore, we have a trade-off between the extraction accuracy and the time variation of the frequencies when choosing $\Delta t$: When $\Delta t$ is too large, the obtained frequency will tend to be the average of its variation over the same period. We chose $\Delta t = 20 \ \text{Myr}$ as the standard FA interval. In some circumstances that require the detection of rapid changes in frequencies, such as the resonance transition, a smaller $\Delta t$ is more favorable (see Sect. \ref{sec:libsack}).
        
        The extracted frequencies are sensitive to the choice of $\Delta t$, and yet their density estimation is relatively robust. Figure \ref{fig:FA} compares the density estimation of the eccentricity frequencies $(g_i)_{1,4}$, which are extracted via FA with three different $\Delta t$ at two different times. The differences are generally small but still notable for $g_2$ and $g_4$ at 150 Myr, and they diminish with time.

            \begin{table*}
        \centering
        \newcolumntype{R}{ >{${}}r<{{}$} }
        \newcolumntype{C}{ >{${}}c<{{}$}}
        \begin{tabular}{C R R C R R C C}
         \hline\hline                        
         \rule{0pt}{1em}
           & \multicolumn{1}{c}{$\mu_0\ [\arcsec / \text{yr}] $  }                     & \multicolumn{1}{c}{ $a\ [\arcsec / \text{yr}]^2 $}          &
           \multicolumn{1}{c}{$b$}          &
           \multicolumn{1}{c}{$\alpha_0 $}             & \mu_1 \ [\arcsec / \text{yr}]   & \sigma_1^2\ [\arcsec / \text{yr}]^2  & A_1\\[0.1em]
        \hline
        \rule{0pt}{1em}
g_1 & 5.759 +  0.006\,T  & 3.37 \cdot 10^{-2\phantom{0}}  & 0.52& -2.25 -  0.50\,T  & &&\\
g_2 & 7.448 -  0.004\,T  & 4.17 \cdot 10^{-4\phantom{0}}  & 0.70& 1.38 +  0.21\,T  & &&\\
g_3 & 17.269 +  0.002\,T  & 6.63 \cdot 10^{-3\phantom{0}}  & 0.43& &&\\
g_4 & 17.896 +  0.005\,T  & 6.88 \cdot 10^{-3\phantom{0}}  & 0.41&  & 17.6755 & 0.0034 & 0.110  -0.012 \,T\\
s_1 & -5.652 - 0.032\,T & 2.68 \cdot 10^{-2\phantom{0}} & 0.83 & 1.12 + 0.16\,T & &&\\
s_2 & -6.709 + 0.030\,T & 1.20 \cdot 10^{-1\phantom{0}} & 0.76 & -2.94 - 1.23\,T & &&\\
s_3 & -18.773 + 0.009\,T & 2.86 \cdot 10^{-2\phantom{0}} & 0.56 & -3.40 - 0.08\,T &  -18.5256 & 0.0028 & 0.023\\
s_4 & -17.707 + 0.013\,T & 1.19 \cdot 10^{-2\phantom{0}} & 0.68 & -1.73 - 0.28\,T & &&\\
\hline
\rule{0pt}{1em}
g_5 & 4.257454-2.1 \cdot 10^{-6}\ T & 4.63 \cdot 10^{-10}  & 0.88\\
g_6 & 28.245226-1.4 \cdot 10^{-4}\ T & 1.40 \cdot 10^{-6\phantom{0}}  & 0.84\\
g_7 & 3.087957-1.2 \cdot 10^{-6}\ T & 4.80 \cdot 10^{-10}  & 1.11\\
g_8 & 0.673024\phantom{\,\ -1.0 \cdot 10^{-2}\ T }& 9.89 \cdot 10^{-11}  & 1.49\\
s_6 & -26.347866+1.5 \cdot 10^{-5}\ T & 1.21 \cdot 10^{-8\phantom{0}}  & 0.85\\
s_7 & -2.992527\phantom{\,\ -1.0 \cdot 10^{-2}\ T }& 8.31 \cdot 10^{-10}  & 1.39\\
s_8 & -0.691737\phantom{\,\ -1.0 \cdot 10^{-2}\ T }& 2.93 \cdot 10^{-11}  & 1.47\\
\hline
\end{tabular}
\caption{Linear and power-law fits for the time evolution of the parameters (Fig. \ref{fig:PDF_fitParams}) of the skew Gaussian mixture model (Eq. \ref{eq:fitModel}) for the fundamental frequencies of the Solar System. Column 1 contains the considered secular frequencies. In Col. 2 we show the linear fits of $\mu_0$ that represent the center of the distribution. The power-law fit of $\sigma^2_0$ has the form $\sigma^2_0(t) = a T^{b}$ (Eq. \ref{eq:powerLaw}), where $T = t/(1 \, \text{Gyr})$ and $a$ and $b$ are given in Cols. 3 and 4, respectively. Linear fits of the skewness parameter $\alpha_0$ are given in Col. 5. The last three columns show linear fits of the secondary mode of $g_4$ and $s_3$ (Eq. \ref{eq:fitModel}).
The parameter $\sigma_0^2$ is fitted from 200 Myr to 5 Gyr in the past, while all the others are fitted from 500 Myr.}      
\label{table:params_fit} 
\end{table*}
    
    \section{Parametric fitting}
    From Sect. \ref{sec:main} we see that the long-term PDFs of the secular frequencies possess a distinct Gaussian-like shape that flattens as time passes. It is interesting to approximate these densities by a simple parametric model, such that its parameters can characterize the shape of the density and summarize its evolution. The model with its fitting parameters can also be used as an approximation of the numerical densities for later application. For this purpose, we used the density estimation of the secular frequencies of the Solar System from the batch $\{Z_{i}\}$, which is composed of 40\,000 different orbits over 5 Gyr.
    The inner fundamental frequencies (i.e., the frequencies of the inner planets, $(g_i,s_i)_{i=1,4}$) are obtained by FA over an interval of 20 Myr. For the outer fundamental frequencies (i.e., the frequencies of the outer planets, $(g_i)_{i=5,8}$, $(s_i)_{i=6,8}$), the FA interval is 50 Myr. The frequency $s_5$ is zero due to the conservation of the total angular momentum of the Solar System. 
    \subsection{Skew Gaussian mixture model} \label{sec:FitInn}
        \cite{laskar2008} found that the 250 Myr-averaged marginal PDFs of the eccentricity and inclination of the inner Solar System planets are described quite accurately by the Rice distribution, which is essentially the distribution of the length of a 2D vector when its individual components follow independent Gaussian distributions. 
        In in our case, the density estimations of the fundamental frequencies of the Solar System resemble Gaussian distributions, but many of them get skewed as time passes; this is especially true for the inner frequencies. To account for this skewness, we propose the skew normal distribution as the fitting distribution to the density estimation of the frequencies:
        \begin{equation} \label{eq:skew}
            f_{\mu_0,\sigma_0,\alpha_0}(x) = \frac{2}{\sigma_0}\phi\left(\frac{x-\mu_0}{\sigma_0}\right)\Phi\left(\alpha \left(\frac{x-\mu_0}{\sigma_0}\right)\right),
        \end{equation}
        where $\alpha$ is the parameter characterizing the skewness, $\phi (x)$ denotes the standard normal probability density distribution with mean $\mu_0$ and standard deviation $\sigma_0$, and $\Phi(x)$ is its cumulative distribution function given by 
        \begin{equation}
            \Phi(x) = \int_{-\infty}^{x} \phi(t)\ dt = \frac{1}{2} \left[ 1 + \operatorname{erf} \left(\frac{x}{\sqrt{2}}\right)\right],
        \end{equation}
        where erf denotes the error function. 
        
        Some of the frequencies, interestingly, have several secondary modes in their density estimation apart from their primary one. Most of the secondary modes, if they exist, are quite small compared to the primary one. They are also often short lived; most of them emerge at the beginning of the diffusion stage and disappear quickly thereafter. Therefore, they are not included in this parametric model, the aim of which is to fit the long-term PDF of the fundamental frequencies.
        However, some persist for a long time and have a small but non-negligible amplitude. To account for these secondary modes, we simply added Gaussian functions to the skew Gaussian distribution and adjusted for their amplitudes so that the fitting distribution is ultimately the skew Gaussian mixture model:
        \begin{equation}
        \label{eq:fitModel}
            f (x) = A_0 f_{\mu_0,\sigma_0,\alpha_0}(x) + \sum_{i=1}^m A_i  \mathcal{N} (\mu_i, \sigma_i^2),
        \end{equation}
        where  $\sum_{i=0}^m A_i= 1$ and m is the number of secondary modes. The secondary modes are much smaller than the primary mode: $A_0 \gg \sum_{i=1}^m A_i$. In our case here, $m=1$ for $g_4$ and $s_3$, while the asymptotic secondary modes for the other frequencies can be considered negligible. 
        
        For the outer fundamental frequencies, we do not observe significant skewness or secondary modes in their density estimations. Therefore, the density estimations of the outer fundamental frequencies can be approximated by a simple Gaussian distribution.
        
        The density estimations of the frequencies are shown at three well-separated times in Fig. \ref{fig:Pdfs_fit}. The diffusion is clearly visible as the density estimations get more and more disperse over time. Moreover, the density estimations get more skewed as time goes by. For $g_3$ and $g_4$, the skewness of the density estimations is small, so we assume $\alpha_0 =0$ (Eq. \ref{eq:skew}) for these frequencies for the sake of simplicity.
        The fundamental frequencies of the outer planets are very stable. Their variations are much smaller than those of the inner frequencies. Taking as an example the most unstable of the outer frequencies, $g_6$, its standard deviation at $-4$ Gyr is only about $2 \cdot 10^{-3}\ \arcsec / \text{yr} $, while that of the most stable inner frequency, $g_2$, is 15 times larger.
        Therefore, when considering a combination involving an inner fundamental frequency and an outer one, the latter can be effectively regarded as a constant.

        The result of our fitting model is also plotted in Fig. \ref{fig:Pdfs_fit}.
        It is remarkable that the density estimations of the frequencies are well approximated by the fitting curve over three different epochs. It should be noticed that the base of our fitting model -- the skew Gaussian distribution -- only has three parameters.  Additional parameters are only needed for some frequencies, for example $g_4$ and $s_3$. Nevertheless, such additional parameters only account for the minor features, and three parameters are sufficient to represent the bulk of the density estimations over a long timescale.
        

    \begin{figure*}[ht!]
    \centering
    \includegraphics[width=17cm]{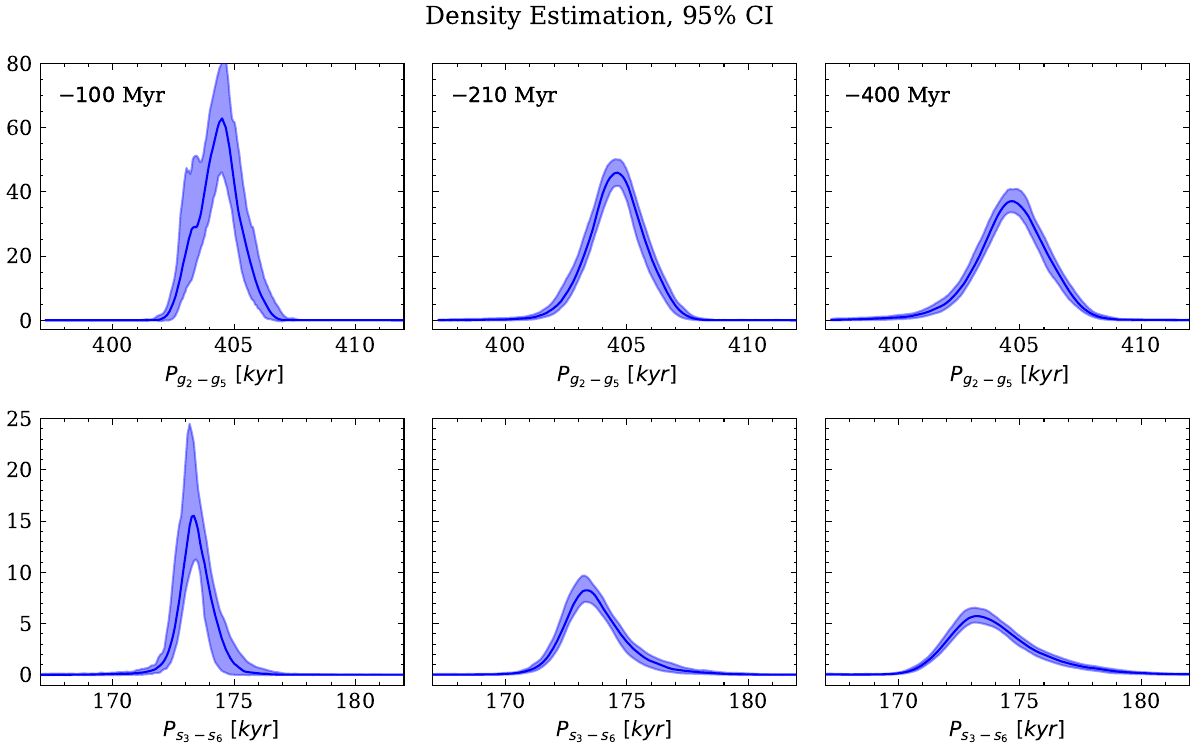}
    \caption{Density estimation with 95\% pointwise CI of the period of $g_2-g_5$ (top row) and $s_3-s_6$ (bottom row) from the combining solutions from $\{X_i\}$, $\{Y_i\}$, and $\{Z_i\}$. Frequencies are obtained by FA over an interval of 20 Myr centered at $-100$ Myr (left column), $-210$ Myr (middle column), and $-400$ Myr (right column). }
    \label{fig:clock}
    \end{figure*}
    \begin{figure*}[ht!]
    \centering        
    \includegraphics[width=18cm]{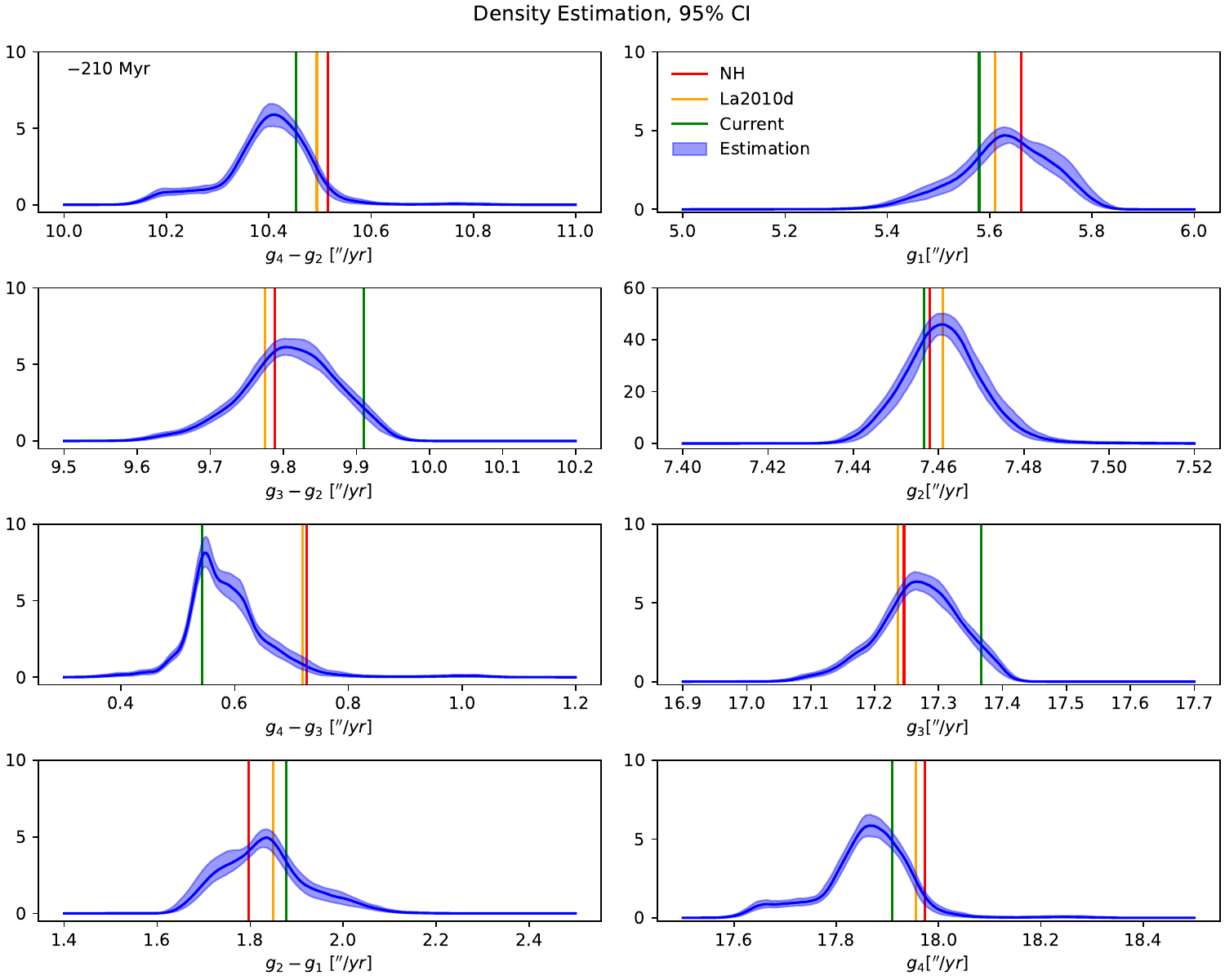}
    \caption{Density estimation with 95\% pointwise CI (blue colors) of the secular frequencies (right column) and their combinations (left column) of the combined solutions from $\{X_i\}$, $\{Y_i\}$, and $\{Z_i\}$, along with the corresponding values extracted from from the La2010d solution (yellow lines), the Newark-Hartford data \citep[red lines;][]{olsen2019}, and the current values (green lines). The frequencies are obtained by FA over an interval of 20 Myr centered at 210 Myr in the past. The time reference is the time of $\{Z_i\}$.}
    \label{fig:NH}
    \end{figure*}
    \begin{figure*}[ht!]
    \centering
     \includegraphics[width=17cm]{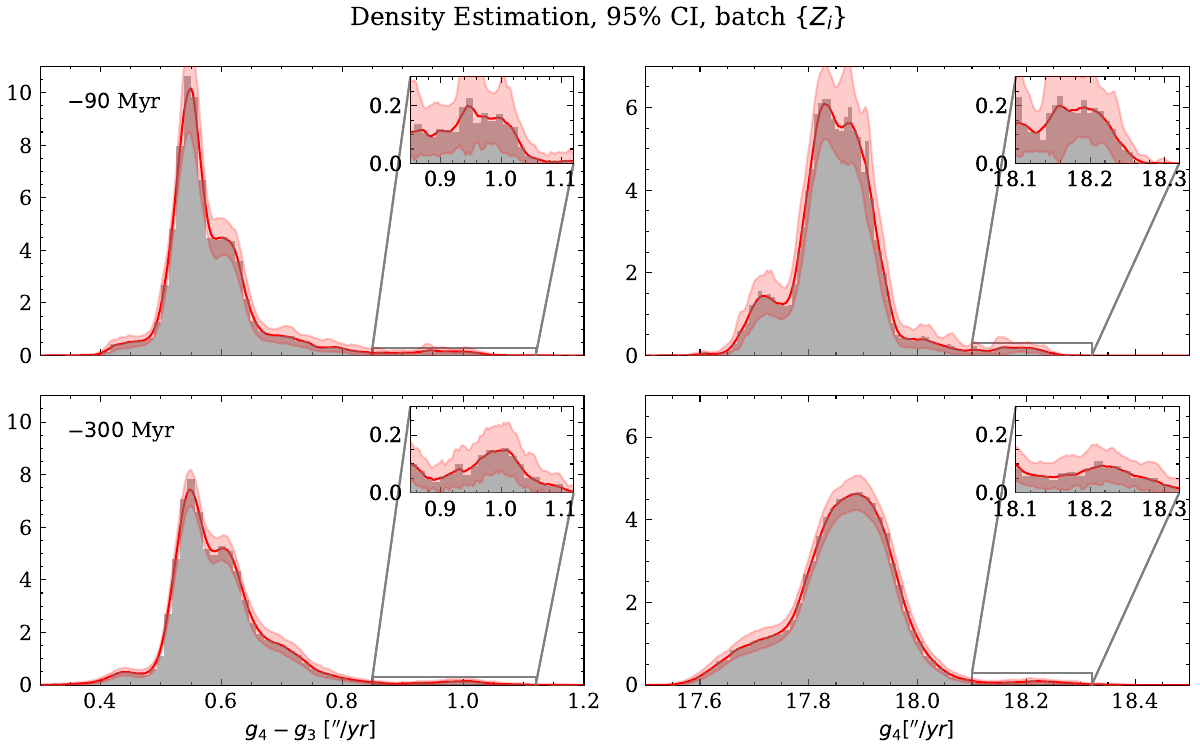}
    \caption{Density estimation with 95\% pointwise CI (red colors) of $g_4-g_3$ (left column) and $g_4$ (right column) at $-90$ Myr (first row) and $-300$ Myr (second row) of $\{ Z_i \}$, overlaid with their histograms (gray blocks) for better visualization. Frequency values are obtained by FA over an interval of 10 Myr. }
    \label{fig:lib}
    \end{figure*}
    \begin{figure*}[ht!]
    \resizebox{\hsize}{!}{\includegraphics{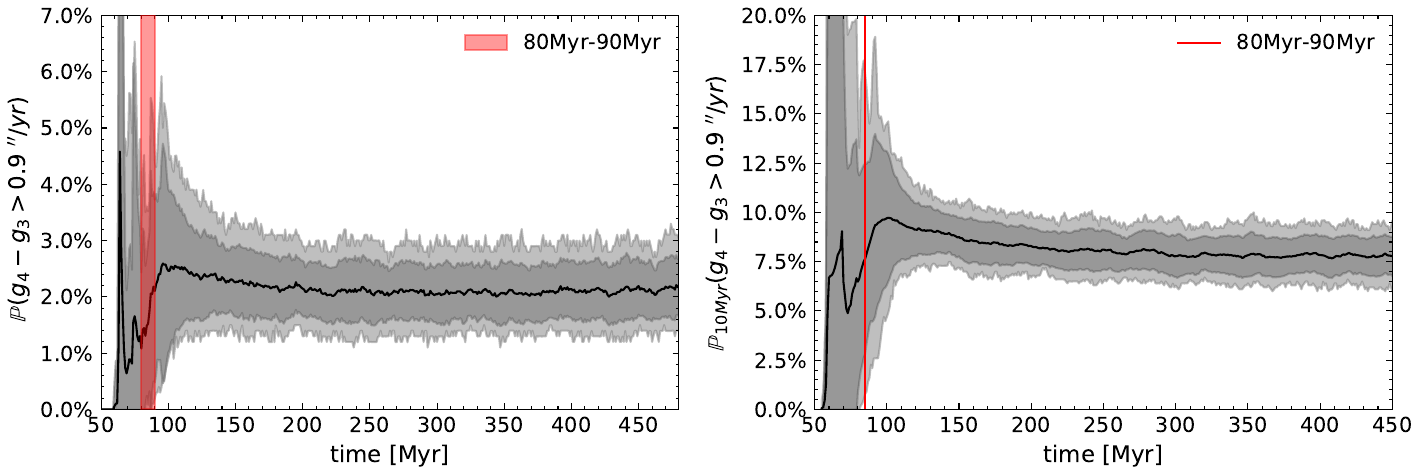}}
    \caption{Percentage of the solutions (black line) whose $g_4-g_3 > 0.9$ at each time (left) and over a 10 Myr interval (right), with their 90\% CI estimation uncertainty (larger gray band) and their standard errors (smaller gray band). The time interval of the Libsack record, which is 80 Myr to $90$ Myr, is denoted by the red band in the left plot and by a red line in the right plot.}
    \label{fig:libPer}
    \end{figure*}

        \subsection{Evolution of the parameters}
        The parameters of our fitting models are extracted by the method of least squares, implemented by the routine $\texttt{curve\_fit}$ in the scipy package in Python.
        To retrieve the statistical distribution (mean and standard deviation) of the parameters of a given model, we implemented a bootstrap approach based on Eq. \eqref{eq:comb}, with the assumption that pointwise standard errors of the KDE estimated by MBB are independent. We remark that, with such an assumption, the variance of the fitting parameters tends to be underestimated. 
        
        The time evolution of the mean of the parameters of the fitting models is shown in Fig. \ref{fig:PDF_fitParams}, along with their $\pm 2$ standard error, for both the past and the future. It turns out that the evolution of $\sigma_0^2$ is robustly fitted from 200 Myr to 5 Gyr in the past by the power-law function 
        \begin{equation} \label{eq:powerLaw}
            \sigma_0^2(t) = a\ T^b,
        \end{equation}
        where $T = t/(1 \, \text{Gyr})$, as shown in Fig. \ref{fig:PDF_fitParams}. For all the other parameters, we performed a linear fit. All these fits are summarized in Table \ref{table:params_fit}.
        
        The differences between past and future evolutions are small and generally tend to decrease with time. Therefore, the fit for the parameters in the past given in Table \ref{table:params_fit} should also be representative of their future evolution. 
        In general, the parameters follow relatively smooth curves with distinct tendencies. The skewness parameters $\alpha_0$, increasing in absolute value, show that the PDFs of the inner fundamental frequencies get more and more skewed over time. The center of the distributions, indicated by  $\mu_0$, does not change significantly compared to $\alpha_0$ and $\sigma_0^2$ (Fig. \ref{fig:PDF_fitParams}). The secondary modes of $g_4$ and $s_3$ are also quite stable.
        
        The diffusion of the frequencies is quantified by the increasing $\sigma_0^2$, which is closely linked to their distribution variance. As the exponents $b$ of the power laws in Table \ref{table:params_fit} are different from unity, the chaotic diffusion of the fundamental frequencies turns out to be an anomalous diffusion process. Interestingly, all the inner frequencies clearly undergo sub-diffusion, that is, the exponent of the power law $b$ is smaller than 1 ($b = 1$ corresponding to Brownian diffusion). Therefore, an extrapolation of the variance of the inner frequencies based on the assumption of linear diffusion over a short time interval would generally lead to its overestimation over longer times. On the contrary, the exponents $b$ are either smaller or larger than unity for the outer frequencies. It should be noted that, because the variations in the outer frequencies are very small, the value of the corresponding exponents $b$ might be overestimated due to the finite precision of FA.

    \section{Geological application}
    The aim of this project is to have a reliable statistical picture of the secular frequencies of the Solar System beyond 60 Myr. It is interesting to put recent geological results in this astronomical framework. First, it can be used as a geological test of our study, and secondly the application provides a glimpse of how astronomical data could be used in a cyclostratigraphy study. We first show how the uncertainty of a widely used dating tool in astrochronology can be quantified, then we apply our work to two recent geological results, which are from the Newark-Hartford data \citep{olsen2019} and the Libsack core \citep{ma2017}. 

    \subsection{Astronomical metronomes}

    Although it is not possible to recover the precise planetary orbital motion beyond 60 Myr, some astronomical forcing components are stable and prominent enough such that they can be used to calibrate geological records in the Mesozoic Era or beyond (see \citealt{laskar2020} for a review). The most widely used is the 405 kyr eccentricity cycle $g_2-g_5$, which is the strongest component of the evolution of Earth's eccentricity. 
    The inclination cycle $s_3-s_6$ has also recently been suggested for the time calibration of stratigraphic sequences \citep{boulila2018, charbonnier2018}. Although $s_3-s_6$ is not the strongest among the obliquity cycles, it is quite isolated from other cycles and thus easy to identify  \citep{laskar2020}.
    The main reason that $g_2-g_5$ and, possibly, $s_3-s_6$ can be used as astronomical metronomes is their stability. Indeed, their uncertainty has to be small to be reliably used for practical application.
    
    In previous work, the uncertainty of the frequency combination was derived from the analysis of a few solutions \citep{laskar2004,laskar2010,laskar2020}. Here we used a much larger number of solutions and the KDE-MBB method to derive both the PDF of the frequencies and their statistical errors. Starting from the results of all the sets of orbital solutions (Sect. \ref{sec:main}), we produced the compound density estimation of the fundamental frequencies and their relevant combinations following the conservative approach in Eq. (\ref{eq:comb}). These data are archived with the paper. 

    With this result, we can reliably estimate the uncertainty of the astronomical metronomes. The uncertainty of the cycles $g_2-g_5$ and $s_3-s_6$ is in fact the density width of the period of these frequency combinations, which is shown in Fig. \ref{fig:clock}. As time goes by, the two metronomes become more uncertain as their density spreads due to the chaotic diffusion, but they are still reliable. At 400 Myr in the past, the relative standard deviations (ratio of standard deviation to mean) of the $g_2-g_5$ and $s_3-s_6$ metronomes are approximately $0.4\%$ and $1.26\%$, respectively.  


    \subsection{Newark-Hartford data}
    In \citet{olsen2019} the astronomical frequencies were retrieved from a long and well-reserved lacustrine deposit. The frequency signals of the Newark-Hartford data are very similar to that of the astronomical solution La2010d, which was taken from 13 available astronomical solutions  \citep{laskar2010}. Having 120\,000 astronomical solutions at hand, we can derive a more precise statistical analysis of this result.
    The fundamental frequencies from the geological record were obtained in the following way.
     The data were dated by the relatively stable 405 kyr cycle ($g_2 - g_5$) and additionally verified by a zircon U–Pb-based age model \citep{olsen2019}. An FA (Sect. \ref{sect:FA}) was performed on the geological data as well as on the eccentricity of the astronomical solution La2010d to retrieve their strongest frequencies. 
    The FA of geological data was cross-checked with that of the astronomical solution to identify its orbital counterpart. For example, the third strongest frequency from the Newark data, $12.989 \ ''/\text{yr}$, was identified with the second strongest frequency from La2010d, $12.978 \ ''/\text{yr,} $ which is $g_3-g_5$. The frequency $g_3$ was then derived by summing the obtained combination with $g_5$, which is nearly constant.
    The same goes for $g_1, g_2$, and $g_4$. Definite values of astronomical fundamental frequencies were thus determined. Unfortunately, the uncertainty estimation of the geological frequencies is not yet available.
    
    Our density estimation of the frequencies was obtained by combining all our samples in the conservative approach outlined in Sect. \ref{sec:combine}. Figure \ref{fig:NH} compares the secular frequencies (right) and their combinations (left) extracted from Newark-Hartford data with our density estimation around the same time in the past. For $(g)_{i=1,3}$, both the geological and La2010d frequencies lie relatively close to the main peak of their corresponding density, while their values for $g_4$ are near the tail. The frequency combinations tell the same story since the geological values involving $g_4$ are off from the main peak. Yet, they are all consistent with our density estimation as there is a non-negligible possibility of finding a secular solution that agrees with the geological data. It should be noted that certain frequencies are significantly correlated, $g_3$ and $g_4$ for example. 
    We cannot assume that they are independent; therefore, we have to calculate the density estimation of their combination directly. 
    
    Given the unavailability of the uncertainty in geological frequencies, the probability of finding the geological frequencies in a numerical orbital solution cannot be obtained directly. However, we can use La2010d, which is the solution from the complete model of the Solar System that matches best with the Newark-Hartford data, as the benchmark for our secular statistics. There are several criteria for determining how good a solution is (i.e., how well it could match with the geological data). A simple and rather straightforward criterion that we used is $\delta = \sqrt{\frac{1}{4}\sum_{i=1}^4 (g_i - g_i^*)^2}$, where $g_i$ and $g_i^*$ are the frequencies from the astronomical solution and the geological data, respectively. A better-suited solution will have smaller $\delta$ and vice versa. We found that in the range from $-200$ Myr to $-220$ Myr, out of the 120\,000 solutions, there are around 5000 (roughly $4.2 \%$ of the total number) that have smaller $s$ than that of La2010d at -210 Myr, which is the value originally used to compare with the geological data.  It should be noted that La2010d is one of 13 available complete solutions. The 95\% CI of the probability of obtaining such a good matching solution from the complete model of the Solar System is thus $(1.37\%, 33.31\% )$ \citep{wilson1927}. Therefore, with the criterion $\delta$, our result is statistically compatible with that of \citet{olsen2019}.

    
    \subsection{Libsack core} \label{sec:libsack}
    \cite{laskar1990,laskar1992b} presented several secular resonances to explain the origin of chaos in the Solar System. In particular, the argument of the resonance $(s_4-s_3) - 2(g_4-g_3)$ is currently in a librational state, that is,
    \begin{equation} \label{res:2:1}
        (s_4-s_3) - 2(g_4-g_3) = 0
    ,\end{equation}
    and moves out to the rotational state around $-50$ Myr. The dynamics can even switch to the librational state of a new resonance:
    \begin{equation} \label{res:1:1}
        (s_4-s_3) - (g_4-g_3) = 0.
    \end{equation}
    This transition corresponds to a change from the 2:1 resonance to the 1:1 resonance of two secular terms, $g_4-g_3$ and $s_4-s_3$.
    

    \cite{ma2017} found a sudden change in the period of a long cycle from 2.4 Myr to 1.2 Myr in the Libsack core of the Cretaceous basin from around $-90$ Myr to $-83$ Myr. This change was also visible in the La2004 astronomical solution, and the long cycle was attributed to the frequency combination $g_4-g_3$, which is visible from the spectrum of the eccentricity of the Earth. Although the exact value before and especially after the transition is not clear, the change in the period is visible from the band power of the core (Fig. 1 of \citealt{ma2017}{}).
    
    This change in $g_4-g_3$ observed in the Libsack core corresponds to a transition from the resonance $(s_4-s_3) - 2(g_4-g_3)$, which is the resonance that the Solar System is currently at, to the resonance $(s_4-s_3) - (g_4-g_3)$. With a large number of astronomical solutions, we could better understand this phenomenon. The transition is usually very fast: The frequency changes quickly to another value and then reverts back just as quickly. Therefore, to study the transition we used a smaller window for the FA, 10 Myr instead  of 20 Myr. 
    Figure \ref{fig:lib} shows the density estimation of $g_4-g_3$ at 90 Myr as well as at 300 Myr in the past. Both have a principle population in the range $[0.4, 0.8 ]\ ''/\text{yr} $ and a small but not insignificant one centered around $1.0\ ''/\text{yr}$, which corresponds to the small chunk of $g_4$ centered at $18.2\ ''/\text{yr}$. The transition observed is a sudden jump in frequency from the main population to the secondary one of $g_4$, and therefore $g_4-g_3$ as well. 
    
    The size of the secondary population is defined as the proportion of the solutions whose $g_4-g_3 > 0.9 \ \arcsec/\text{yr}$ and is denoted as $\mathbb{P}(g_4-g_3>0.9\ ''/\text{yr})$. The rate of transition is defined as the proportion of the solutions whose $g_4-g_3 > 0.9\ ''/\text{yr}$ over 10 Myr and is denoted as $\mathbb{P}_{10 Myr}(g_4-g_3>0.9\ \arcsec /\text{yr})$. Both are shown in Fig. \ref{fig:libPer}. During the predictable period, that is, from now until $-50$ Myr, no transition is observed. After $-50$ Myr, a transition can occur; its rate rises until 100 Myr, when the percentage of a secondary population stays relatively stable at around $2.1 \% \pm 0.5 \% $ at a time, and the rate of the transition could also be determined to be $ 8 \% \pm 1 \% $ every 10 Myr during this period.
    At 80-90 Myr, when the transition was detected in the Libsack core, that rate of transition during this period is found with our numerical solutions to be $7.7 \% \pm 4.5 \% $.

    \section{Conclusion}  
    In this work we give a statistical description of the evolution of the fundamental frequencies of the Solar System beyond 60 Myr, that is, beyond the predictability horizon of the planetary motion, with the aim to quantify the uncertainty induced by its chaotic behavior. The base of our analysis is 120\,000 orbital solutions of the secular model of the Solar System. The PDF of the frequencies is estimated by the KDE, and its uncertainty is evaluated by the MBB; both methods are tested via numerical experiments.  
    
     We benchmarked the secular model by sampling the initial conditions in different ways and then compared the density estimation of their solutions with one another and finally with the complete model. The results are twofold. First, regardless of how initial conditions were sampled, their density estimation will converge toward a single PDF; after this overlap, a robust estimation is guaranteed.
     Secondly, the density estimation of the secular model is compatible with that of the complete model of the Solar System. This agreement means that the results of the secular model, with superior computational simplicity, can be used for application to geological data. 
     
     We observe that the density estimations of the fundamental frequencies can be well fitted by skew Gaussian mixture models. The time evolution of the parameters $\sigma_0^2$, related to the frequency variances, follows power-law functions. Interestingly for the inner fundamental frequencies, the exponents of such power laws are all smaller than 1, which indicates that they undergo sub-diffusion processes. 
     
     We show several examples of how this result can be used for geological applications. First, the uncertainty of any astronomical frequency signal is fully quantified, so that, for example, a proper quantitative response can be given to the question of how stable the astronomical metronomes are. 
     With this statistical framework, previous results from geological records beyond 60 Myr can also be interpreted with a more comprehensive approach. A more quantitative answer, not only about the possibility but also about the probability of the occurrence of an astronomical signal in geological data, can be made. 
     Apart from these direct applications, a more systematic approach could make full use of the density estimation of frequencies. The method TimeOpt from \cite{meyers2018}, for example, shows that it is possible to combine the uncertainty from astronomical signals with geological records to derive an effective constraint for both.  In fact, any similar Bayesian method could use the density estimation of frequencies as proper priors. 
     
    \begin{acknowledgements}
    NH is supported by the PhD scholarship of CFM Foundation for Research. This project has received funding from the European Research Council (ERC) under the European Union’s Horizon 2020 research and innovation programme (Advanced Grant AstroGeo-885250) and from the French Agence Nationale de la Recherche (ANR) (Grant  AstroMeso ANR-19-CE31-0002-01).
    \end{acknowledgements}
      

     


    
    \bibliographystyle{aa} 
    \bibliography{CDFFSS}
    \begin{appendix}
    \end{appendix}
\end{document}